%% file: main.tex
\RequirePackage{silence}
\documentclass[%
 aip,
 cha,
 amsmath,amssymb,
 reprint,%
 superscriptaddress,%
 noeprint,
longbibliography, %
]{revtex4-2}

\input{preamble}

\makeatletter
\def\@email#1#2{%
 \endgroup
 \patchcmd{\titleblock@produce}
 {\frontmatter@RRAPformat}
 {\frontmatter@RRAPformat{\produce@RRAP{*#1\href{mailto:#2}{#2}}}\frontmatter@RRAPformat}
 {}{}
}%
\makeatother
\begin{document}

\preprint{AIP/123-QED}


\title{Creation and Microscopic Origins of Single-Photon Emitters in Transition Metal Dichalcogenides and Hexagonal Boron Nitride}

\author{Amedeo Carbone}
\affiliation{Department of Electrical and Photonics Engineering, Technical University of Denmark, 2800 Kgs. Lyngby, Denmark}
\affiliation{NanoPhoton - Center for Nanophotonics, Technical University of Denmark, 2800 Kgs. Lyngby, Denmark}
\affiliation{Walter Schottky Institute, Technical University of Munich, Am Coulombwall 4a, 85748 Garching, Germany}

\author{Diane-Pernille Bendixen-Fernex de Mongex}
\affiliation{Nanolab - National Centre for Nano Fabrication and Characterization, Technical University of Denmark, 2800 Kgs. Lyngby, Denmark}

\author{Arkady V. Krasheninnikov}
\affiliation{Institute of Ion Beam Physics and Materials Research, Helmholtz-Zentrum Dresden-Rossendorf, 01328 Dresden, Germany}

\author{Martijn Wubs}
\affiliation{Department of Electrical and Photonics Engineering, Technical University of Denmark, 2800 Kgs. Lyngby, Denmark}
\affiliation{NanoPhoton - Center for Nanophotonics, Technical University of Denmark, 2800 Kgs. Lyngby, Denmark}

\author{Alexander Huck}
\affiliation{Center for Macroscopic Quantum States (bigQ), Department of Physics, Technical University of Denmark, 2800 Kgs. Lyngby, Denmark}

\author{Thomas W. Hansen}
\affiliation{Nanolab - National Centre for Nano Fabrication and Characterization, Technical University of Denmark, 2800 Kgs. Lyngby, Denmark}

\author{Alexander W. Holleitner}
\email{holleitner@wsi.tum.de}
\affiliation{Walter Schottky Institute, Technical University of Munich, Am Coulombwall 4a, 85748 Garching, Germany}%
\affiliation{Munich Center for Quantum Science and Technology (MCQST), Schellingstr. 4, 80799 Munich, Germany}
\affiliation{TUM International Graduate School of Science and Engineering (IGSSE), Boltzmannstr. 17, 85748 Garching, Germany}

\author{Nicolas Stenger}
\email{niste@dtu.dk}
\affiliation{Department of Electrical and Photonics Engineering, Technical University of Denmark, 2800 Kgs. Lyngby, Denmark}%
\affiliation{NanoPhoton - Center for Nanophotonics, Technical University of Denmark, 2800 Kgs. Lyngby, Denmark}

\author{Christoph Kastl}
\email{christoph.kastl@wsi.tum.de}
\affiliation{Walter Schottky Institute, Technical University of Munich, Am Coulombwall 4a, 85748 Garching, Germany}%
\affiliation{Munich Center for Quantum Science and Technology (MCQST), Schellingstr. 4, 80799 Munich, Germany}
\affiliation{TUM International Graduate School of Science and Engineering (IGSSE), Boltzmannstr. 17, 85748 Garching, Germany}

\date{\today}

\begin{abstract}
We highlight recent advances in the controlled creation of single-photon emitters in van der Waals materials and in the understanding of their atomistic origin. We focus on quantum emitters created in monolayer transition-metal dichalcogenide semiconductors, which provide spectrally sharp single-photon emission at cryogenic temperatures, and the ones in insulating hBN, which provide bright and stable single-photon emission up to room temperature. After introducing the different classes of quantum emitters in terms of band-structure properties, we review the defect creation methods based on electron and ion irradiation as well as local strain engineering and plasma treatments. A main focus of the review is put on discussing the microscopic origin of the quantum emitters as revealed by various experimental platforms, including optical and scanning probe methods.  
\end{abstract}

\maketitle

\section*{Introduction}

On-chip sources of single photons are important building blocks for applications in quantum communication, computing, simulation, and metrology \cite{Rodt2021, Couteau2023, Montblanch2023}. Single-photon emitters (SPEs) in layered bulk and two-dimensional (2D) van der Waals (vdW) materials, originally discovered in atomically thin semiconductors \cite{Tonndorf2015, Chakraborty2015, Koperski2015, He2015, Srivastava2015, Tonndorf2017}, insulators\cite{Tran2016}, and vdW heterostructures\cite{Baek2020}, are being explored as an emerging solid-state platform \cite{Barthelmi2020, Azzam2021, Turunen2022, MichaelisdeVasconcellos2022, Gupta2023, Kaul2025}, which is complementary to established technologies, such as semiconductor quantum dots\cite{Hepp2019} or color centers in diamond\cite{Atature2018}. Layered vdW materials promise not only ease of hybrid integration but also opportunities for monolithic integration \cite{Turunen2022, Kianinia2022} into larger photonic circuits, for example waveguides \cite{Tonndorf2017, Peyskens2019, Li2021}, cavities \cite{Froch2022,Iff2021}, microlenses \cite{Preuss2023}, detectors \cite{Najafi2015} or electronic circuits\cite{Loh2024a}. In initial studies, SPEs were randomly discovered at the edges or folds of crystals\cite{Tonndorf2015, Koperski2015, Srivastava2015, Chakraborty2015, He2015} suggesting that mechanical strain plays a major role in their creation or activation, not only making them distinct from other solid state SPEs, but also offering the possibility to create and tune their properties via strain fields \cite{Iff2019, Grosso2017, Mendelson2020, Patel2024}. Recent advances in methods for creating quantum emitters, for example
using irradiation \cite{Krasheninnikov2007, Klein2019, Ziegler2019, Fournier2021, Gale2022, Zhang2022a, Parto2021, Cianci2023, Howarth2024, Micevic2022} and strain engineering \cite{Palacios-Berraquero2016, Kumar2015, Kern2016, Branny2017, Cai2018, Brooks2018, Luo2018, Luo2019, Wang2021, Sortino2021, Yu2021, Zhao2021, VonHelversen2023, Chen2024, Zhao2025}, have opened a route for scalable, nanoscale integration of SPEs with unparalleled positioning accuracy\cite{Palacios-Berraquero2017, Kern2016, Branny2017, Mitterreiter2020, Parto2021, Gale2022}.

\begin{figure*}[tbhp]
  \centering
  \includegraphics[width=1.9\columnwidth]{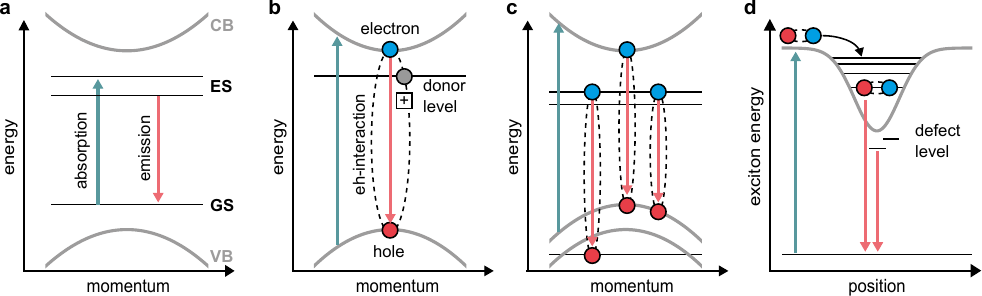}
  \caption{\textbf{Quantum emission from defects.} \textbf{a} Color centers derived from deep in-gap states with ground state (GS) and multiple excited states (ES). The color center is directly addressable in absorption or emission. \textbf{b} Shallow donors or acceptors states localize excitons into bound complexes by mutual electron-hole interactions. \textbf{c} Defect-localized excitons can hybridize both to deep in-gap states and to shallow extended band states. \textbf{d} Local strains or moiré potentials trap excitons. Point defects may facilitate radiative recombination and single-photon emission.}
  \label{fig:defect_models}
\end{figure*}

To date, the scientific debate about the atomistic origin of single-photon emission in vdW materials has not been conclusively settled for many of the experimentally observed emission lines, despite a decade of extensive research. Defects are commonly assumed to play an essential role in enabling single-photon emission\cite{Linhart2019, Thompson2020, Kumar2015, Linhart2019, Moody2018, Mitterreiter2021, Xu2022}, and their atomic structure can be inferred, in particular for monolayer transition metal dichalcogenides (TMDCs), from atomic resolution microscopy, such as scanning tunneling microscopy (STM), non-contact atomic force microscopy (AFM), or aberration corrected scanning transmission electron microscopy (STEM) \cite{schuler2020electrically, Mitterreiter2021, Mitterreiter2020, Park2021, Singla2024,Sohn2024, Qiu2024, Krivanek2010}, as well as from atomistic calculations based on \textit{ab-initio} approaches, such as density functional theory (DFT) \cite{Gupta2019, Refaely-Abramson2018, Gao2021a, Amit2022, Bertoldo2022, Kirchhoff2024, Maciaszek2022}. Nevertheless, establishing a direct link between the atomic structure and the optical signature of defects has proven to be difficult. In particular, the microscopic nature of quantum emitters present in hexagonal boron nitride (hBN) is still largely under debate, despite many attempts to compare \textit{ab-initio} calculations with experiments~\cite{Mendelson2021, Aharonovich2022, Fischer2021, Fischer2023, Zhigulin2023a, Hayee2020, Singla2024}. From a theoretical point of view, predicting the optical properties of defects in a periodic crystal is very demanding. To describe excitonic effects \textit{ab-initio}, solving the Bethe-Salpeter equation (BSE) is required with a so-called $GW$ correction on top of the electronic structure calculated within the framework of the ‘conventional’ DFT with local or semi-local exchange and correlation functionals. For defects or heterostructures, these calculations require rather large supercells \cite{Bertoldo2022, Kundu2023}, which in turn means that large computational resources are needed to converge the exciton energies to about 10 meV - 100 meV \cite{Attaccalite2011, Refaely-Abramson2018, Gao2021a, Kirchhoff2024}. Exciton-phonon coupling, which is especially relevant for room-temperature emitters in hBN, can be included using effective models\cite{Wigger2019, Fischer2023}, although full \textit{ab-initio} methods are emerging as well\cite{Paleari2019, Chen2020, Chan2024}. Additionally, many optically active defects or defect complexes exhibit charged states resulting in structural modification via, for example, the Jahn-Teller-effect~\cite{Watanabe2009, Gupta2019, Xiang2024, Tan2020, Jansen2024}, which at present are difficult to properly account for in \emph{ab-initio} calculations. Lastly, while the emission spectrum from a single defect is easily detectable, its optical absorption spectrum, which is usually the quantity that \emph{ab-initio} calculations predict, is more difficult to access experimentally. In this respect, photoluminescence excitation (PLE) spectroscopy is a suitable tool to gain insights into absorption properties, excited states, and vibronic structure of single defects~\cite{Grosso2020, Preuss2022, Fischer2023}. Cavity-enhanced absorption spectroscopy of He ion-irradiated MoS\textsubscript{2} reported quantitative measurements of defect-related sub-gap absorption down to $10^{12}$ defects/cm$^2$, which still translates into about $10^3$ defects probed within a typical laser spot \cite{Sigger2022}. By contrast, early reports in GaAs achieved already single-defect sensitivity: cross sections as low as $\sigma \approx \qty{1e-18}{cm^{-2}}$ could be measured by monitoring current (fluctuations) in vertical tunneling structures upon sub-gap excitation~\cite{Snow1993}, similar to recent photocurrent experiments on MoS$_2$/hBN/graphene and strained WSe$_2$ heterostructures~\cite{Paur2020, Hotger2023a}.

When aiming to correlate atomic structure and optical functionality, the mismatch in length scales between the actual structure (few \AA ngstroms) and the spatial resolution in a typical far-field optical experiment (hundreds of nanometers) presents a formidable experimental challenge. Statistical photon counting provides robust, yet not perfect, evidence whether a given emission line stems from a single physical object~\cite{Fishman2023}. It is not perfect, because different atomic defects or multiple emission lines with varying emission power may be present within the typical size of a far-field optical detection spot. In this context, optical near-field~\cite{Darlington2020, Yanev2024, Niehues2025} or cathodoluminescence (CL) spectroscopy~\cite{Shevitski2019,Hayee2020} are powerful techniques to investigate quantum emitters in 2D materials beyond the diffraction limit, enabling, for example, the correlation of local strain fields and single-photon emission. Yet, a direct experimental link between the atomic structure of defects, defect-associated electronic states and their optical emission spectrum will ultimately be necessary to unambiguously establish their atomistic structure-function relationships~\cite{robinson2021engineering}.

Here, we review recent advances in the controlled creation of SPEs in vdW materials and in the understanding of their atomistic origin. We focus on quantum emitters created by irradiation engineering in direct band gap TMDC semiconductors, which provide spectrally sharp single-photon emission at cryogenic temperatures, and in insulating hBN, which provide bright and stable single-photon emission up to room temperature. Due to the accelerated expansion of this research field in the past years, we do not aim for completeness. 
For the present work, we discuss selected representative examples that emphasize irradiation engineering and microscopic origin of single-photon emission. Additionally, we refer the reader to other reviews focusing on different aspects of quantum emitters in van der Waals materials, such as integration into photonic structures \cite{Kianinia2022, Azzam2021}, electrical control and optoelectronic integration into quantum light-emitting devices \cite{Loh2024a}, tunability of single photon emission via strain and electric fields \cite{Yu2024}, magneto-optical properties, engineering and coherence of spin defects\cite{Vaidya2023, Fang2024}, integration of 2D material-based quantum emitters into photonic platforms, with a focus on deterministic placement, coupling efficiency, and device scalability for quantum technologies \cite{Montblanch2023}, the role of strain engineering \cite{Hou2025}, exciton dynamics and excitonic quantum light generation in vdW heterostructures \cite{Niehues2024}, in-depth analysis of intrinsic and extrinsic defects in hBN \cite{Im2025}, and computational discovery of defect-based quantum emitters \cite{Bassett2019, Ping2021}.
The review is structured as follows. To begin with, we introduce different classes of optically active emitters based on localized exciton states emphasizing the roles of deep vs. shallow states, different exciton binding energies, and varying local potentials in the exciton localization and single-photon emission process (\secref{defects}). We discuss the fundamentals of irradiation engineering of lattice defects in thin vdW materials (\secref{fundamentals}). Next, we review selected examples of quantum emitters as generated by local strain engineering, where vdW materials provide opportunities to create and tailor SPEs in ways not possible in traditional bulk material systems (\secref{strain}). Plasma treatment provides a simple means to introduce optically active defects or defect complexes near the material surface on a large scale (\secref{plasma}). Ion irradiation (\secref{ion}) is commonly exploited for the creation of positioned SPEs, either based on the local introduction of single point defects into the host crystal or on the local patterning/sputtering of the host material. Although ion implantation is a powerful workhorse for color center creation in bulk materials, only few works to date discuss the ion implantation of functional impurities into 2D and quasi-2D vdW materials (\secref{implantation}). By contrast, electron beam irradiation (\secref{e-beam}) has attracted significant interest as a simple approach for site-selective creation or activation of single-photon emitters. As an example of non-charged particle irradiation, \secref{xray} highlights focused X-ray and UV-light irradiation as a tool to both introduce and spectroscopically interrogate defect states in 2D semiconductors. 
In the last part, we emphasize the importance of carefully identifying the atomistic origin of SPEs both in hBN and TMDCs (\secref{nature}). \secref{atomistic} highlights recent advances towards the correlation of optical properties and structure at the atomic scale, and in an outlook, we address current and future challenges of SPEs in 2D vdW materials (\secref{outlook}).

\section{\label{sec:defects}Classes of optically active defects}

In hBN, both single photon emitters and spin-defects are mostly associated with deep-level color centers. In TMDCs, however, multiple mechanisms, such as defect-bound excitons and strain localization are relevant, often simultaneously, as discussed in the following. Color centers are typically related to states inside the band gap due to vacancies, impurities, or complexes thereof that give rise to significant sub-gap absorption, thereby changing the apparent color of the crystal (hence the name color center). If the in-gap states are located sufficiently deep inside the gap, the optical properties of the defect can be understood in an atomic-like picture (\figref[a]{defect_models}). To first order, interactions with electronic states in the conduction and valence bands can be neglected in this case because of the separation of energy scales (on the order of eV) and because of the localization of the defect states (on the order of a unit cell or less). The localized electrons form a ground state (GS) and several possible excited states (ES), whereby the electron configuration is dictated by group theoretical considerations and exclusion principles according to the symmetries of the defect and the host crystal~\cite{Ivady2020a, Abdi2018}. Direct excitation of the defect into one of the higher-lying excited states can be achieved by efficient absorption of photons at an energy below the optical gap of the host crystal. Single-photon emission then typically occurs from one of the lower-lying excited states after intra-defect relaxation processes and according to the symmetry-allowed dipole selection rules~\cite{Abdi2018}. Single photon emission can often be observable up to room temperature in wide-gap materials, including hBN. The energy of the absorption and emission lines can be significantly (on the order of eV) below the energy of the band gap of the host material. Additionally, color centers often have a spin-dependent fine structure that allows control of spin transitions with suitable radio-frequency or microwave fields. This enables optically detected magnetic resonance measurements~\cite{Stern2022, Gottscholl2020, Chejanovsky2021, Guo2023} or coherent excitation and read-out of spin states, e.g. for the {\VB} defect in hBN~\cite{Gottscholl2021, Rizzato2023}. In general, color centers interact very efficiently with local vibronic excitations (or local phonon modes), in particular for heavy impurities or defects with large structural distortions. For hBN, this often results in complex emission spectra with multiple series of phonon side bands below the zero-phonon line~\cite{Khatri2019, Wigger2019, Feldman2019, Fischer2023}.

By contrast, \figref[b]{defect_models} illustrates the case of a shallow impurity level, such as a donor (or acceptor) in 2D semiconducting TMDCs. A bound electron-hole pair (or exciton) can localize to a shallow donor site via the mutual Coulomb interaction between the electron in the conduction band (blue), the hole in the valence (red), the donor electron (grey), and last but not least the screened core of the donor impurity (square). In the simplest model, the interaction is described in an effective mass approximation, implying that the formed particle complex delocalizes sufficiently over multiple unit cells~\cite{Klingshirn2012}. In contrast to color centers, direct excitation of bound-exciton states via sub-gap absorption is inefficient, although it is in principle detectable down to the limit of a single bound exciton~\cite{Snow1993}. Rather, free excitons are excited by band-to-band transitions, and at low temperatures they will subsequently localize to the impurity sites via diffusion processes. In low-temperature luminescence spectra, bound excitons manifest as sharp emission lines at energies of approximately 10-100 meV below the free exciton emission. D$^0$X denotes excitons bound to neutral donors, while D$^+$X denotes excitons bound to ionized donors, leaving behind the positively charged core. Along these lines, the notation for acceptor bound states is A$^0$X (neutral) and A$^-$X (charged), although the latter is often considered energetically unstable~\cite{Klingshirn2012}. For traditional bulk semiconductors, such as III-V materials, the binding energy scales as $E^\text{b}_\text{D$^+$X} < E^\text{b}_\text{D$^0$X} < E^\text{b}_\text{A$^0$X}$. For 2D vdW semiconductors, Ref.~\onlinecite{Mostaani2017} provides extensive calculations covering monolayer MoS\textsubscript{2}, WS\textsubscript{2}, MoSe\textsubscript{2}, and WSe\textsubscript{2}. Note that, in general, the binding energy $E^\text{b}$ of the defect-bound complex is not equivalent to the energy difference $\Delta E$ between the free exciton emission and the defect-bound exciton emission, because radiative recombination and dissociation may not necessarily result in the same final state. For example, Ref.~\onlinecite{Mostaani2017} calculates for D$^+$X in monolayer MoSe\textsubscript{2} $\Delta E = \qty{100}{meV}$ and $E^\text{b}_\text{D$^+$X} = \qty{6.5}{meV}$. This distinction resolves the apparent contradiction between the experimentally observed large detuning of the emission energy between defect-bound and free excitons, which can be on the order of 200 meV, and the small thermal activation energy of the defect emission (i.e. the binding energy of the complex), which is typically on the order of 10 meV \cite{Carozo2017}. Furthermore, these values exemplify that single-photon emission from excitons bound to shallow impurities, as demonstrated for example in Ref.~\onlinecite{Loh2024} for dilute Nb impurities in WS\textsubscript{2}, necessarily requires low temperatures, because otherwise free excitons will not localize onto the defect site.

In 2D TMDCs, the reduced dielectric screening results in large exciton binding energies on the order of a few hundred meV~\cite{Kylanpaa2015a}. For carriers localized onto a defect site, the large exciton binding energy complicates a clear distinction between shallow and deep levels, since excitonic states can mix contributions from different levels over a large energy scale. In this case, \emph{ab-initio} calculations may provide model-free insights into the optical properties of the defects. For instance, \figref[c]{defect_models} sketches the level scheme of MoS\textsubscript{2} with sulfur-vacancy defects, as calculated by DFT with GW corrections~\cite{Mitterreiter2021}, comprising both occupied quasi-resonant defect levels overlapping with the spin-orbit split valence band and unoccupied in-gap states below the conduction band. In this case, BSE calculations predict a manifold of possible exciton states with a varying admixture of tightly localized defect versus more extended band states, as depicted conceptually in \figref[c]{defect_models}, and with transition energies up to \qty{700}{meV} below the free exciton~\cite{Amit2022}. 

Lastly, spatial variations of the free exciton energy due to strain fields\cite{Rosati2021} and variations of the local dielectric environment\cite{BenMhenni2025} play an important role in exciton localization and corresponding single-photon emission in the case of 2D TMDCs. Local strain potentials may induce effective potential wells for excitons at mesoscopic length scales~\cite{Zhao2023} with typical confinement energies on the order of a few meV due to renormalization of both band gap and exciton binding energy with strain~\cite{Rosati2021, Glazov2022}. These strain fields can effectively guide single excitons into quantized, quantum-dot-like states at low temperatures, which in turn emit single photons upon radiative recombination \figref[d]{defect_models}. Although the microscopic origin of the single-photon emission has not been unambiguously resolved, both experimental and theoretical evidence suggests that atomic point defects play an important role in the recombination process. For example in monolayer WSe\textsubscript{2}, calculations predict that strain locally shifts the exciton states into resonance with in-gap defect states resulting in inter-valley hybridization, efficient filling, and subsequent radiative decay from the defect state~\cite{Linhart2019, Branny2017}. Note that moiré potentials in heterostructure can localize excitons at similar length and energy scales with a corresponding single-photon emission at low temperatures~\cite{Baek2020}.

\section{Fundamentals of Irradiation Engineering}
\label{sec:fundamentals}

Irradiation with energetic particles, such as ions or electrons, is one of the most powerful and established techniques to tune material properties by introducing defects or impurities in a controllable manner~\cite{Hoflich2023, Krasheninnikov-2010, Li2017, Schleberger-2018, Krasheninnikov-2010, Zhao-2019, Telkhozhayeva2024}. Depending on the energy and type of the particle, as well as on the morphology and electronic properties of the irradiated material, different mechanisms of energy transfer from the projectile to the target are responsible for the formation of atomic defects (\figref[a, b] {fundamentals}). Furthermore, the evolution of the system after impact of the particle, which includes any subsequent annealing and interaction with the environment, also strongly affects the final resulting types and concentrations of defects. For 2D materials, interaction with the environment can substantially change defect types and/or their concentration, as any unsaturated bonds at the surface tend to become passivated, for example by picking up other atoms, molecules, or radicals from the environment~\cite{Zan2012rek}. 

For electron irradiation, the three main mechanisms for defect creation are the knock-on (or ballistic) damage, inelastic interactions due to electronic excitations and ionization, and beam-induced chemical etching, as illustrated in \figref [c-e]{fundamentals}.~\cite{Susi2019_NatRevPhys, Egerton2013_Ultramicroscopy, Krasheninnikov2007, Zhao2017} Moreover, defects can also appear due to simultaneous contributions of several of these mechanisms.~\cite{Kretschmer2020, Lehnert} The Coulomb interaction between the incoming electrons and the atomic nuclei results in elastic energy transfer from the electron to the recoil atom (\figref[c]{fundamentals}). The minimum electron energy required to cause displacement of atoms is referred to as the electron knock-on displacement threshold energy. The latter is related through the relativistic binary collision formula to the recoil atom displacement threshold energy, which is the minimum kinetic energy the recoiling atom should acquire to leave its position without immediate recombination. It depends both on the mass of the recoil atom and its chemical binding energy. At typical electron energies (60-300 keV) used in TEM, the energy transfer can be high enough for knock-on damage. With regard to semiconducting and insulating 2D materials, such as TMDCs and hBN, formation of defects has also been reported~\cite{Bui2023, Kretschmer2020, Speckmann2023, Harriet2022, Dash2023, Dash2025, Lin2015_NatComm} at electron energies well below the knock-on threshold, which indicates that other damage mechanisms are active as well. For 2D MoS\textsubscript{2}, a combination of electronic excitations and elastic scattering can in principle give rise to the displacement of atoms at rather low electron energies~\cite{Dash2023}. Electronic excitations/ionizations (\figref[d]{fundamentals}) alone cannot result in the formation of defects in an ideal crystal with high electron mobility due to a quick delocalization of the excitation~\cite{Kretschmer2020}. Different physical processes contribute to inelastic energy transfer, including electronic excitations, ionization of target atoms, or electron–phonon coupling \cite{Ziegler1985}. Beam-induced etching mechanisms (\figref[e]{fundamentals}), are very relevant to atomically thin targets, despite their comparably weak interaction with the electron beam, and they are active both in hBN~\cite{Bui2023a} and TMDCs~\cite{Harriet2022, Dash2023}. For example, degradation of 2D MoTe\textsubscript{2} in an oxygen partial pressure above 10$^{-7}$ torr was shown by \emph{in-situ} TEM, while MoS\textsubscript{2} was found to be inert to electron irradiation under oxygen environment on the time scales of the experiment~\cite{Harriet2022}.
Most likely, etching is the dominant channel of defect production under the electron beam in the SEM (typical energy is 1-30 keV), although charge accumulation in the material or the substrate followed by Coulomb explosion can also give rise to the appearance of defects~\cite{Wei2013_ACSNano}. Furthermore, adatoms present on the surface can also enhance the cross section for ballistic damage~\cite{Jain2024}, as new bonds formed between the material atoms and adatoms can decrease local in-plane bonding. For MoTe\textsubscript{2}, the presence of hydrocarbons accelerated the etching by up to a factor of forty, which was attributed to break-up and oxygen release from the complexes~\cite{Harriet2022}. Break-up of hydrocarbon molecules, which can be ubiquitous in the TEM or SEM column, also gives rise to the deposition of amorphous carbon or other organic materials on top of the sample~\cite{Yagodkin2022, Dyck2022}. Deposition can proceed very fast at low electron energies, as the cross section for the break-up of the molecules increases with decreasing electron energy~\cite{Huth2012}. 

For ion irradiation~\cite{Nastasi1996, Hoflich2023}, nuclear stopping dominates in the case of relatively slow and heavy ions (\figref[a]{fundamentals}), and it originates from collisions and corresponding energy transfer between the ion and the nuclei of the atoms in the target governed by screened Coulomb interactions. By contrast, electronic stopping is a result of inelastic collisions between the moving ion and the electrons in the target, and it is important at high ion energies and/or for light ions. The crossover between nuclear and electron stopping depends on the ion mass. For hydrogen ions (protons) and helium ions, electronic stopping always dominates. Yet, defect production can still be caused through knock-on damage, as the amount of energy deposited through excitations may not be sufficient to displace atoms, especially in highly conductive systems.

\begin{figure}[tbhp]
  \centering
\includegraphics[width=\columnwidth]{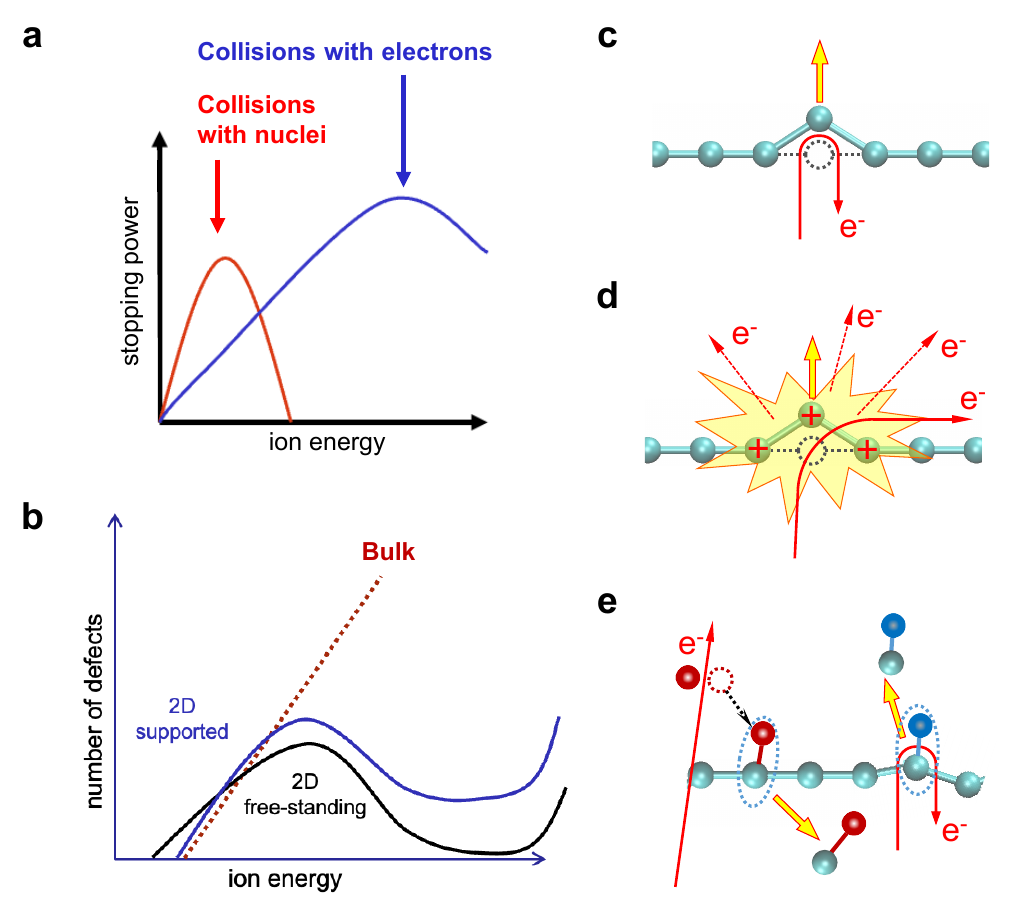}
  \caption{
  \textbf{Fundamentals of the irradiation-mediated engineering of 2D materials.} \textbf{a} Schematic illustration of energy loss per unit length (stopping power) of the ion, with the main loss mechanisms being dependent on ion energy. \textbf{b} Sketch of the number of defects produced in a bulk system, a free-standing, and a supported 2D material by energetic ions as a function of ion energy. \textbf{c} Knock-on damage production mechanism upon electron irradiation, e.g. in the TEM. \textbf{d} Excitation/ionization damage production mechanism. \textbf{e} Beam-induced chemical etching and adatom-mediated damage production upon electron irradiation.}
  
  
  
  \label{fig:fundamentals}
\end{figure}

Plasma treatment can be considered as a low-energy ion irradiation. Plasmas typically achieve fluences much larger than those achieved by irradiation with ion beams, with the drawback that control over the fluence of ions and their energy distribution is limited. Impurities can be introduced into 2D materials directly by ion implantation~\cite{Bangert2013_NL, Susi2017, Tripathi2018, Bui2022, Villarreal2024, Willke2015, Telychko2022} or foreign atoms can fill vacancies created by energetic particles~\cite{Wang2012nl, Ma2014acsn}. 

The reduced dimensionality of 2D materials affects their response to ion bombardment (\figref[b] {fundamentals}). Due to the planar geometry of the target and its thinness, the ion loses only a fraction of its energy, and the development of collisional cascades is suppressed. As a result, the number of defects produced in few-layer 2D materials will first increase with ion energy above a certain threshold, but will then drop at higher energies due to a smaller probability for the ion to displace an atom~\cite{Krasheninnikov2020}. Especially for free-standing sheets, higher-energy ions should go through the 2D system without producing much damage~\cite{Ghaderzadeh2021}. Defects in 2D materials can appear again when the electronic stopping power is high enough to locally ionize and "melt" the material \cite{Akcoltekin2008}. This stands in contrast to bulk systems, where the total number of defects is proportional to ion energy, but defects appear deeper in the sample with increasing ion energy. The situation may be different for supported 2D materials. At low ion energies (below \qty{100}{eV}), the substrate should decrease the amount of damage created by the energetic ions. On the other hand, for ions with medium energies (for example He and Ne ions with energies above \qty{10}{keV} in a helium ion microscope), the production of defects in the 2D material can be governed by the backscattered ions and atoms sputtered from the substrate rather than by direct ion impact~\cite{Kretschmer2018, Mitterreiter2020, Villarreal2024}. For implantation into 2D materials, the ion energy should be relatively low (a few tens or at most hundreds of eVs), much lower than what is used for ion implantation into bulk materials.

\section{Strain Engineering}\label{sec:strain}

\begin{figure*}[thb]
  \centering
  \includegraphics[width=1.85\columnwidth]{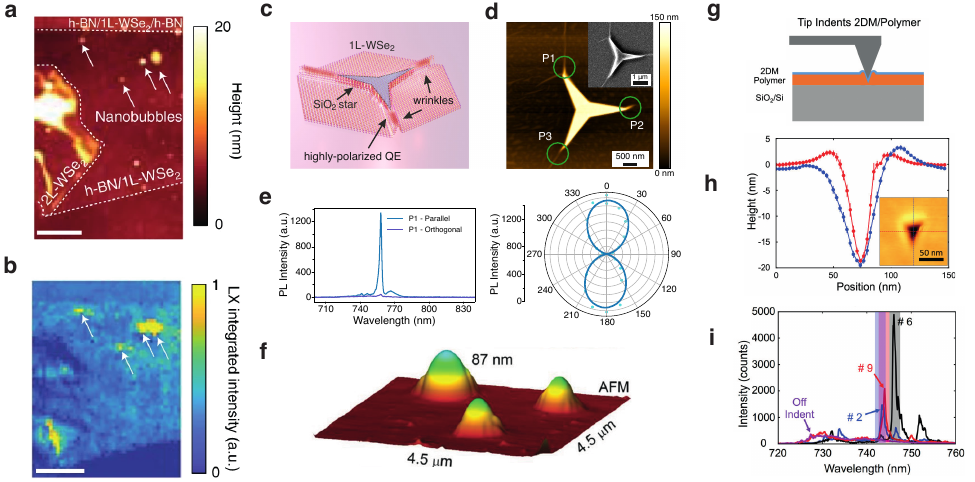}
  \caption{\textbf{Strain engineering of SPEs.} \textbf{a} AFM topography and \textbf{b} integrated near-field PL (range 1.5–1.6 eV) image of exfoliated monolayer $\mathrm{WSe_2}$ on top of hBN. The white arrows indicate that PL emission is spatially localized on nanobubbles sites. All scale bars, 500 nm. Reproduced with permission from \citet{Darlington2020}, Nat. Nanotechnol. \textbf{15}, 854–860 (2020). Copyright 2020 Springer Nature. \textbf{c} Sketch and \textbf{d} AFM image (inset: SEM image) of monolayer $\mathrm{WSe_2}$ on a a star-shaped nanopillar. \textbf{e} (left) In-plane polarization-resolved PL spectrum and (right) corresponding polar plot (degree of polarization about 99\%) taken from the P1 region in \textbf{d}. Reproduced with permission from \citet{Paralikis2024}, npj 2D Mater. Appl. \textbf{8}, 59 (2024). Copyright 2024 Authors, licensed under a Creative Commons Attribution Non Commercial License 4.0 (CC BY-NC). \textbf{f} AFM image of a region of a bulk $\mathrm{WS_2}$ flake treated with $\mathrm{H^+}$-irradiation, showing formation of nanodomes. Reproduced with permission from \citet{Tedeschi2019}, Adv. Mater. \textbf{31}, 1903795 (2019). Copyright 2019 WILEY‐VCH Verlag GmbH \& Co. KGaA, Weinheim. \textbf{g} Working principle of "quantum calligraphy": an AFM tip applies sufficient load to plastically deform a polymer underneath a 2D material, leaving a strain configuration. \textbf{h} Line profile of the strain configuration in \textbf{g}. \textbf{i} PL spectra with SPEs generated by the local strain in \textbf{g}. Reproduced with permission on from \citet{Rosenberger2019} ACS Nano \textbf{13} (1), 904-912 (2019). Copyright 2019 American Chemical Society.}
  \label{fig:strain}
\end{figure*}

We start by reviewing briefly key works on strain-induced localization of quantum emitters in 2D materials since they represent the first demonstration of site-selective creation of quantum emitters in the field. Strained 2D materials are also commonly used in combination with irradiation engineering as discussed in the following sections. Local variations of strain have been discussed as the origin for quantum light emission in TMDCs since their first observations~\cite{He2015, Chakraborty2015, Tonndorf2015, Srivastava2015, Koperski2015, Kumar2015, Kern2016, Palacios-Berraquero2017, Branny2017}. Strain variations often occur naturally due to wrinkles, folding, or bubbles beneath the layer. Inhomogeneous strain profiles from such corrugations facilitate the mesoscopic confinement of single excitons due to modifications of both band gap and exciton binding energy \cite{Niehues2024}. Upon radiative recombination, the trapped excitons thus create single photons at locations predefined by the nano-strain. In particular, nanobubbles with sizes on the order of 10-100 nm can form during the dry stamping process for the transfer of TMDCs~\cite{Shepard2017}. Using near-field optical microscopy and hyperspectral imaging, the formation of doughnut-shaped trapping potentials and their correlation to localized exciton emission was clearly demonstrated for such nanobubbles in monolayer WSe\textsubscript{2}~\cite{Darlington2020} (\figref[a-b]{strain}). Through heating of a poly(methyl methacrylate) membrane (PMMA) in the TMDC transfer process, hydrocarbon-filled bubbles with sub-µm sizes can be formed in large quantities underneath TMDC layers with universal strain profile independent of the bubble size. Choosing the substrate material, the strain profile can be tuned, and with this the emission wavelength, as shown for MoS\textsubscript{2} and a range of substrate materials~\cite{Tyurnina2019}. Strain in TMDCs can also be introduced deliberately and site-selectively by patterning of the substrates with nanopillars and subsequent positioning of (few-layer) TMDCs onto the nano-corrugated surface \cite{Palacios-Berraquero2017, Branny2017, Palacios-Berraquero2017, Mukherjee2020, Cai2018, Iff2018, Luo2018, Iff2021, Luo2019, Wang2021, Sortino2021, Chen2024, Chhaperwal2024, Singh2025, Yu2025}. Although initially demonstrated for intralayer excitons in direct-gap monolayer TMDCs, pillar-driven strain engineering of SPEs can be extended to interlayer excitons in 2D heterostructures, such as MoSe\textsubscript{2}/WSe\textsubscript{2} heterobilayers\cite{Zhao2023a}  or WSe\textsubscript{2} homobilayers \cite{Ripin2023}. Single photon emission from 2D heterolayers not only extends the tunability of the materials platform, but the large out-of-plane dipole moment of interlayer (localized) exciton transitions enables also electrical control of both emission energy and coupling to phonons \cite{Ripin2023}. Alternative methods for introducing local corrugations include the use of nanoparticles \cite{VonHelversen2023}, nanorods \cite{Kern2016}, nanowires \cite{Cai2017, So2021}, nano-antennas \cite{Azzam2023}, waveguides \cite{Errando-Herranz2021}, etched holes (for both WSe\textsubscript{2}~\cite{Kumar2015} and MoSe\textsubscript{2} \cite{Yu2021}), and trenches \cite{Li2022}, as well as strain tuning via application of an electrostatic field on a suspended monolayer \cite{Wu2025b}. 
By applying strain on MoTe$_2$ via nano-pillar arrays, \citet{Zhao2021} achieved single-photon emission up to \qty{1550}{nm} (telecom C-band). It has been also shown that the emission wavelength of quantum light emitters localized on nanopillars can be tuned by applying an electric field with a few-layer graphene electrode \cite{Mukherjee2020}. Tuning of the emission wavelength can be also achieved via electrostatically actuated microcantilevers in SPEs as defined by utilizing nano-pyramid patterns~\cite{Kim2019} or via piezo-electric strain tuning along wrinkles~\cite{Iff2019}. Very recently, \citet{Wu2025a} exploited valley-spin locking in pillar-strained monolayer WSe$_2$ using a cross-circular polarization scheme to consistently reduce the background emission from free excitons and exciton complexes, thus achieving single photon emission with improved purity.

Wrinkles in a hBN-TMDC bilayer structure placed on a patterned substrate have also been shown to be an optimal source of strain~\cite{Daveau2020}, while strain induced with Au nanoparticles exhibits single-photon emission with temperature-dependent energy loss~\cite{VonHelversen2023}. Similarly, Au tips ~\cite{Peng2020} and star-shaped Au particles~\cite{Daveau2020} were applied for WSe\textsubscript{2}. Yet another approach for local strain creation is based on a chemomechanical modification~\cite{Utama2023}, achieved with non-covalent surface modification with aryl diazonium chemistry.

The polarization and angular momentum of the emitted single photons can be defined by the symmetry of the local strain field (\figref[c-e]{strain}), using, for example, gold chiral nanoparticles~\cite{Lee2024} or patterned strain-mediating structures with elongated star-shaped structures~\cite{Paralikis2024}. Exploring the optical properties of the pattern-forming material together with their structure opens possibilities for modifying the dynamics of the emitter and the far-field radiation pattern. Radiative emission can thus be directly coupled to waveguide modes~\cite{Errando-Herranz2021}, local cavity resonances~\cite{Iff2021}, or efficiently scattered to the far field with dielectric nano-antennas~\cite{Sortino2021, Azzam2023}. 

Nanodomes with highly symmetric radial strain profiles can be generated controllably via $\mathrm{H^+}$ irradiation (see \figref[f]{strain}). After penetration of the first layers, $\mathrm{H^+}$-ions form molecular hydrogen in the first interlayer. TMDC material and dome size thus provide means to control the strain profile and the emission wavelength~\cite{Tedeschi2019}, while capping with an hBN layer prevents dome deflation~\cite{Cianci2023}. Yet another approach for the site-controlled domes formation is the irradiation of polymer-supported vdW materials with low-energy electrons, where hydrogen is formed in the electron beam-induced degradation of the polymer~\cite{Pandey2024}. Lastly, nanometer-sized indentations carefully patterned with an atomic force microscope in a soft substrate, such as a spin-coated polymer layer, is another highly reproducible approach for exciton confinement via local strain modification~\cite{Rosenberger2019, Abramov2023, So2021a, Chuang2024, Lee2024a} (see \figref[g-i]{strain}).

For hBN, SPEs were created by silica nanopillars \cite{Proscia2018}, nano-indentation \cite{Karankova2025}, and on nanoscale wrinkle sites \cite{Yim2020}. Also in this case, strong photoluminescence (PL) with good in-plane polarization and photon purity was observed. Furthermore, SPE creation and enhancement at the same time was achieved by employing gold-coated silicon pillars with an alumina spacer \cite{Sakib2024}, yielding a 10-fold PL enhancement and 2.46-fold reduction in lifetime. Notably, \citet{Chen2025} employed Ag nanowire/hBN/graphene tunnel junctions to electrically drive strain-activated defect emission, although without reporting evidence of single-photon emission.

Depending on the type of defect and the band structure of the host material, strain may have different consequences on light emission. In semiconducting TMDCs, strain can funnel excitons which are subsequently trapped and recombine radiatively, often mediated by atomic defect states. By contrast, in wide band gap hBN, strain creates local electric fields and shifts the energy levels of intrinsic defects, thus allowing to modulate and fine-tune the optical emission spectrum. Furthermore, strain may also have an important impact on the defect creation process. In this context, first-principle calculations, for both hBN and TMDCs, predict that bi-axial strain changes the formation energy of defect structures depending on the relative size of the impurity atom~\cite{Santra2024}.

\section{Plasma Treatment}\label{sec:plasma}

\begin{figure*}[thb]
  \centering
  \includegraphics[width=1.75\columnwidth]{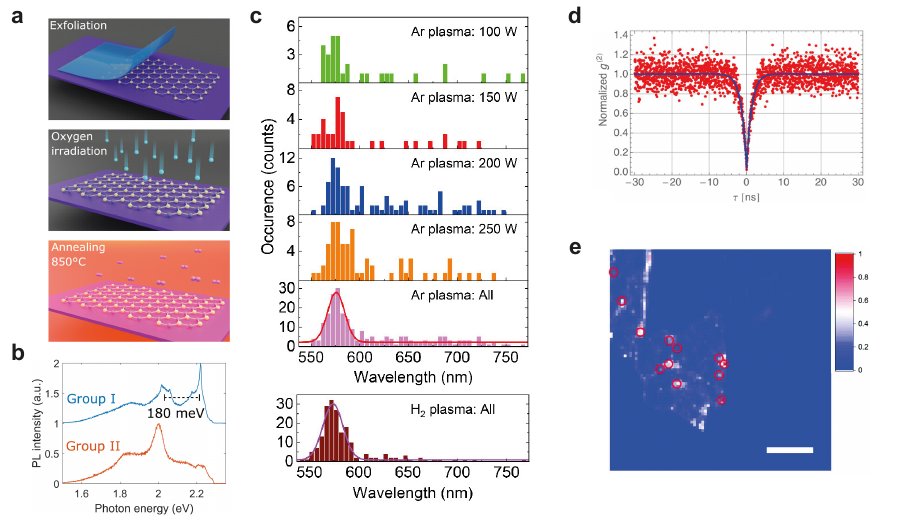}
  \caption{\textbf{SPEs in hBN as fabricated by plasma treatment.} \textbf{a} Typical fabrication steps for creating SPEs in exfoliated hBN (top) based on an $\text{O}_2$-plasma treatment (middle) and subsequent high-temperature annealing in controlled atmosphere (bottom). \textbf{b} The luminescence of $\text{O}_2$-plasma treated hBN typically exhibits several lines, which can be grouped into distinct lineshape categories (here group I and II as in Ref.~\onlinecite{Fischer2021}) and which are red-shifted by $\sim$160 - 200~meV with respect to the zero-phonon line of the SPEs. Reprinted from \citet{Fischer2021}, Science Advances
7, 7138–7155 (2021). Copyright 2021 Authors, licensed under a Creative Commons Attribution Non Commercial License 4.0 (CC BY-NC) License. \textbf{c} Statistics of the zero-phonon line of emitters in hBN as treated with an $\text{Ar}$-plasma at certain plasma powers (top) and with a $\text{H}_2$-plasma (bottom). Reprinted with permission from \citet{Zeng2024}, ACS Applied Materials \& Interfaces \textbf{16}, 24899–24907 (2024). Copyright 2024 American Chemical Society. \textbf{d} Normalized second-order autocorrelation of an emitter in hBN fabricated by $\text{O}_2$-plasma treatment with $g^{(2)}(0) $ = 0.0034 (47). Reprinted with permission from \citet{Vogl2018}, ACS Photonics \textbf{5}, 2305–2312 (2018). Copyright 2018 American Chemical Society. \textbf{e} Luminescence map of SPEs (red circles) in hBN fabricated by an $\text{Ar}$-plasma with a characteristic random spatial distribution. Scale bar: 10~µm. Reprinted with permission from \citet{Xu2018}, Nanoscale \textbf{10}, 7957–7965 (2018). Copyright 2018 Royal Society of Chemistry.}
  \label{fig:plasma}
\end{figure*}

There have been several reports on quantum emitter fabrication in hBN by plasma application, including Ar-~\cite{Xu2018, Zeng2024}, H\textsubscript{2}-~\cite{Zeng2024}, or O\textsubscript{2}-plasmas~\cite{Vogl2018, Fischer2023}. Typically, thin layers of hBN are mechanically exfoliated onto a substrate and then brought in contact to one of the mentioned plasmas to induce defects, while the activation of single-photon emission relies on a subsequent high-temperature thermal annealing (\figref[a]{plasma}). Most reports have utilized an annealing temperature in the range of 750 to \qty{900}{\degreeCelsius} for several minutes either in Ar-gas\cite{Xu2018, Vogl2018}, N\textsubscript{2}-gas\cite{Fischer2021} or even air \cite{Zeng2024} to suppress the fluorescence background in the photoluminescence spectra and to enhance the overall emitter stability. The characteristics of the corresponding quantum emitters vary both in spectral fingerprints (\figref[b]{plasma}) and in possible microscopic origins~\cite{Fischer2023} (cf. \secref{nature}). Moreover, the energy of the zero-phonon line typically varies both with the type of plasma used and with the applied duration and power (\figref[c]{plasma})~\cite{Vogl2018,Zeng2024}. Nevertheless, plasma-induced quantum emitters in hBN have proven to exhibit very good quantum optical properties (\figref[d]{plasma}), and can be easily generated across a rather large spatial area of the exposed materials (\figref[e]{plasma})~\cite{Vogl2018,Xu2018,Fischer2021, Fischer2023,Zeng2024}. For TMDCs, there are several reports that Ar \cite{Chow2015, Wu2017} or N\textsubscript{2} \cite{Qian2022a} plasma exposure can generate luminescent defects as well. However, for all studies, the reported spectra are consistent with ensemble emission, and single-photon emission was not demonstrated.

\section{Ion Irradiation}\label{sec:ion}

\begin{figure*}[htbp]
  \centering
  \includegraphics[width=1.95\columnwidth]{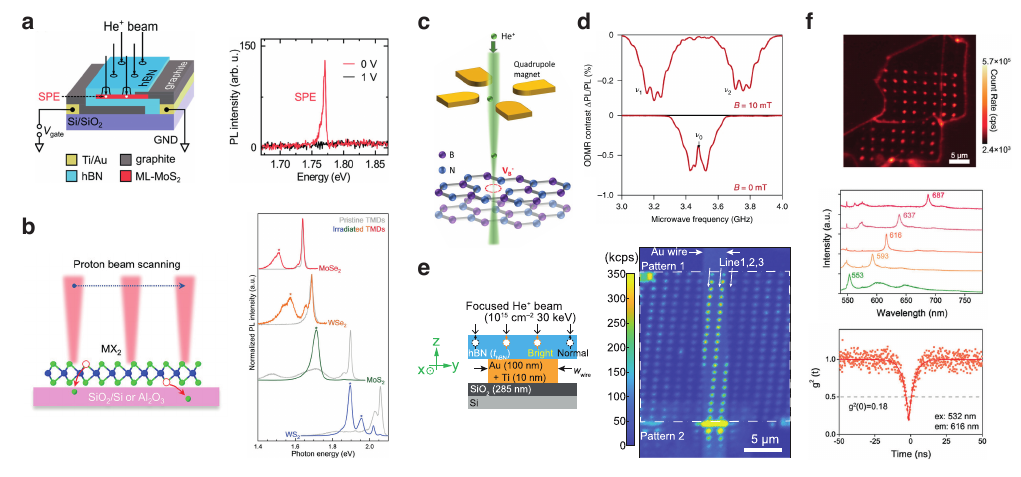}
  \caption{\textbf{Creation of SPEs utilizing ion irradiation.} \textbf{a} With the focused ion beam, defects can be produced in TMDCs embedded in device structures. Reproduced with permission from \citet{Hotger2021}, Nano Lett. \textbf{21} (2), 1040–1046 (2021). Copyright 2021, American Chemical Society. \textbf{b} A proton beam creates chalcogen vacancies in monolayer semiconductors. Reproduced with permission from \citet{Zhang2022a}, Adv. Optical Mater. \textbf{10}, 2201350 (2022). Copyright 2022 Wiley‐VCH GmbH. \textbf{c} Sketch of the highly focused ion beam that knocks out boron atoms in hBN. Reproduced with permission from \citet{Liang2023}, Adv. Optical Mater. \textbf{11}, 2201941 (2023). Copyright 2022 Wiley‐VCH GmbH. \textbf{d} ODMR spectra of defects in hBN at vanishing and at finite magnetic fields. Reproduced with permission from \citet{Gottscholl2020}, Nat. Mater. 19, 540–545 (2020). Copyright 2020, under exclusive license to Springer Nature Limited. \textbf{e} Plasmonic enhancement of photoluminescence of defects in hBN created by a focused He{$^{+}$} ion beam. Reproduced from \citet{Sasaki2023}, Appl. Phys. Lett. \textbf{122}, 244003 (2023), with the permission of AIP Publishing. \textbf{f} Photoluminescence of an ion-irradiated array in hBN after thermal annealing, with associated PL spectra and second-order correlation functions. Reproduced with permission from \citet{Liu2023}, Adv. Optical Mater. \textbf{12}, 2302083 (2024). Copyright 2023 Wiley‐VCH GmbH.}
  \label{fig:irradiation}
\end{figure*}

TMDC monolayers can be locally irradiated by a focused beam of $\rm He^{+}$ ions to introduce defects, for example by the help of a scanning helium ion microscope (HIM). The local crystal modification can act as a spatial trap for excitons in TMDC monolayers, as reported in a series of papers for different materials: MoS\textsubscript{2}~\cite{Klein2018, Klein2019, Barthelmi2020, Mitterreiter2021, Mitterreiter2020, Klein2021a, Barthelmi2024}, WS\textsubscript{2}~\cite{Micevic2022}, WSe\textsubscript{2}~\cite{Howarth2024}, MoSe\textsubscript{2}~\cite{Iberi2016}. In Ref.~\onlinecite{Klein2018}, it was shown how the introduced defects impact the vibrational, optical, and spin-valley properties in MoS\textsubscript{2} supported on a silicon oxide substrate. hBN encapsulation prior to irradiation revealed for the first time optically active defects with spectrally narrow emission lines near \qty{1.75}{eV}, that is approximately \qty{200}{meV} below the optical band gap, which were later confirmed to be SPEs due to defects created at controlled positions in the MoS\textsubscript{2} layer~\cite{Klein2019}. The emitters show long-term stability over multiple cooling cycles~\cite{Barthelmi2020}, although inhomogeneous broadening between different emitters exceeds the natural linewidth~\cite{Klein2019}. Individual defect sites positioned with an accuracy below \qty{10}{nm} were observed with scanning tunneling microscopy, and pristine and oxygen-passivated sulphur vacancies were identified as the most abundant types of defect created by He-ion beam irradiation~\cite{Mitterreiter2020}, in agreement with estimates from SRIM simulations. The optical emission energies are consistent with pristine sulphur vacancies in monolayer MoS\textsubscript{2}~\cite{Mitterreiter2021}, which are predicted to form localized exciton states, hybridized to extended band states (cf. \figref[c]{defect_models}). Spin characteristics, magnetic field-dependent photoluminescence, and dipolar emission patterns of these quantum emitters were studied in Refs.~\onlinecite{Hotger2021, Barthelmi2024}. A distinct advantage of the focused $\rm He^{+}$ ion beam is that it allows creation of single-photon emitters at desired positions after encapsulation of the monolayer and after device fabrication~\cite{Hotger2021} (\figref[a]{irradiation}). Helium ion irradiation has also been used to modify other monolayer TMDCs\cite{Iberi2016}. In monolayer WS\textsubscript{2} irradiated with a HIM, narrow photoluminescence peaks with energies below the exciton energies were identified as local defect states~\cite{Micevic2022}, but no single-photon statistics have been demonstrated yet. 

Other ion sources were explored as well. Focused $\rm Ga^{+}$ ion beam exposure of monolayer WS\textsubscript{2} resulted in a strong modification of the photoluminescence even at locations far away (\qty{>10}{\micro\meter}) from the exposure sites due to redeposition of Ga ions~\cite{Sarcan2023}, suggesting that the heavy Ga ions are poorly suited for the generation of single defect sites in TMDCs. \citet{Zhang2022a} systematically studied the impact of proton exposure on the excitonic properties of all the archetypical direct-gap monolayer semiconductors, namely MoS\textsubscript{2}, WS\textsubscript{2}, MoSe\textsubscript{2}, and WSe\textsubscript{2} (\figref[b]{irradiation}). In all cases, emission lines 100-200 meV below the free exciton with lifetimes beyond \qty{1}{ns} were found, whereby the sulphides generally showed two orders of magnitude brighter defect emission than the selenides. As the underlying origin, chalcogen monovacancies were tentatively proposed, based on defect statistics gained from high-resolution STEM. A follow-up study elaborated in detail on the gate dependence of proton-induced emitters in MoSe\textsubscript{2} and suggested bound excitons (cf. \figref[b]{defect_models}) localized to neutral and negatively charged Se vacancies as the origin of the defect emission~\cite{Chen2023b}. Note that none of the above studies experimentally demonstrated single-photon emission from the ion-induced defect state, but rather reported spectra consistent with optical emission from defect ensembles.

We now turn to irradiation studies on hBN. An important difference is that TMDCs are typically studied as monolayers, while defect emission from hBN crystals is most commonly studied in multilayer, bulk-like samples. Quite commonly, irradiated hBN samples exhibit broad luminescence peaks around 1.5 eV, which are attributed to charged boron vacancies ({\VB}). Although so far only ensemble luminescence has been reported from these defects, their spin triplet (S = 1) ground state makes them promising for quantum sensing applications based on optically detected magnetic resonance (ODMR) schemes~\cite{Healey2023, Vaidya2023}.
Intriguingly, {\VB} defects can also be used to manipulate nearby nuclear spins in hBN using the hyperfine interaction between nuclear spins and {\VB} electron spins~\cite{Gao2022}, similar to what has been done with NV centers in diamond~\cite{Schwartz2018}. The {\VB} defects can be created by many different methods, including irradiation (see \figref[c]{irradiation}) with protons~\cite{Hennessey2024, Froch2021}, helium ions~\cite{Grosso2017, Liang2023, Sarkar2024, Gao2021, Gao2022, Ren2023, Ren2024, Guo2022, Sasaki2023, Zhou2024, Liu2023, Carbone2025}, nitrogen ions~\cite{Clua-Provost2023, Clua-Provost2024, Guo2022, Kianinia2020, Healey2023, Sortino2024}, gallium ions~\cite{Gottscholl2020, Mu2022}, carbon ions~\cite{Guo2022, Baber2022}, oxygen plasma~\cite{Vogl2018, Fischer2021}, xenon ions~\cite{Kianinia2020, Glushkov2022}, argon ions~\cite{Guo2022, Kianinia2020, Ren2024}, or neutrons~\cite{Toledo2018, Gottscholl2020,Gottscholl2021, Gottscholl2021b, Kumar2022}. Alternative techniques that do not directly involve ion beams are neutron-driven nuclear transmutation doping~\cite{Durand2023, Udvarhelyi2023, Clua-Provost2024a, Mu2025}, transmutation doping~\cite{Li2021}, femtosecond laser writing~\cite{Gao2021femto} and AFM-induced damage~\cite{Xu2021}. \citet{Gottscholl2020} first reported broad defect emission centered around \qty{800}{nm} (\qty{1.5}{eV}) after irradiation with various particles (neutrons, lithium and gallium ions). Careful analysis of ODMR experiments (\figref[d]{irradiation}) identified the luminescent defects as {\VB} (cf. \secref{nature}).

For sensing applications, the creation of spin defects within a few atomic layers from the material surface may be advantageous~\cite{Kianinia2020}. In this context, low-energy ${\rm He}^{+}$ ion implantation (\qty{200}{eV} -- \qty{3}{keV}) was used to create shallow {\VB} defects integrated into gold plasmonic structures producing enhanced ODMR contrast (up to 50\%) and enhanced brightness~\cite{Gao2021}. \citet{Sasaki2023} report local magnetic-field sensing with a hBN quantum sensor nanoarray, based on {\VB} defects created by helium irradiation directly on microfabricated gold wires (\figref[e]{irradiation}). The magnetic field created by current flow in the wires was resolved by an ODMR detection scheme with irradiated defect patches as small as \qty{100}{nm^2}, thus demonstrating the potential for high spatial resolution imaging.
Towards optimizing the creation process for {\VB}, \citet{Suzuki2023} specifically compared high-temperature irradiation and post-irradiation annealing, with optimal ODMR contrast obtained at \qty{500}{\degreeCelsius} and \qty{500}{\degreeCelsius} -- \qty{600}{\degreeCelsius}, respectively. Lastly, ion irradiation can also create {\VB} defects in hBN crystals with engineered nanophotonic environments~\cite{Gao2021, Mendelson2022, Xu2023, Qian2022c}. {\VB} defects within nanophotonic bullseye structures revealed spectrally tunable narrowing of the otherwise broad photoluminescence spectrum \cite{Froch2021}. Similarly, {\VB} defects were coupled to bound states in the continuum using a nanostructured hBN as a dielectric metasurface with spectral selectivity \cite{Do2024, Sortino2024}. Experiments on {\VB} monolithically integrated into hBN nanobeam cavities provided evidence that the zero-phonon line of the {\VB} defect is located at \qty{773}{nm}, which is otherwise obscured by strong inhomogeneous and phonon broadening of emission spectra~\cite{Qian2022c}.

For {\VB}, an important consideration is the efficiency of the creation process, especially when its charged nature is taken into account. \citet{Guo2022} estimated a lower bound for the generation probability around $10^{-3}$\% from the photon count pumped by a pulsed laser. Recently, \citet{Gong2023} proposed to extract the {\VB} density by comparing the spin coherence time $T_2$ measured with two different schemes (XY-8 and DROID). Under the assumption that the splitting of the $m_s=\pm 1$ spin sublevels results predominantly from the interaction of {\VB} with the local electric field generated by nearby, randomly distributed charges (see also \secref{nature}), rather than from local strain effects, the defect density can be estimated from comparison of coherence time and splitting to numerical simulation. Support for the splitting-by-local-electric-fields hypothesis comes from a recent theoretical study~\cite{Udvarhelyi2023} and from measurements showing that the $E$ splitting vanishes in few-layer hBN, that is when such local electric fields are almost absent~\cite{Durand2023}. To date, the microscopic origin of the charge environment is not fully understood. Commonly, the charge density derived from such a model is assumed to be equal to the number of charged {\VB} defects. However, a recent work suggested that a phenomenological background charge should also be taken into account to improve the quantitative estimation of the {\VB} density created upon irradiation \cite{Carbone2025}.

Next, we focus on the irradiation engineering of other types of luminescent centers, where single-photon emission was demonstrated. In one of the first studies, from 2017~\cite{Grosso2017}, a 100-nm thick hBN flake was irradiated by $\rm He^{+}$ ions in a He-ion focused ion beam (FIB) and then post-annealed in argon atmosphere at \qty{1000}{\degreeCelsius}. Due to a reduction in background fluorescence, which was present in the as-prepared samples in the irradiated regions, SPEs near \qty{2}{eV} could be resolved, and ZPL tuning via strain up to \qty{6}{meV} was demonstrated. The importance of annealing is evident in Ref.~\onlinecite{Liu2023}, where focused $\rm He^{+}$ irradiation creates sites that display broad PL spectra, but no single-photon emission. Only after subsequent annealing to \qty{1050}{\degree C} in an oxygen atmosphere, the exposed sites showed single-photon emission with energies ranging from \qty{1.8}{eV} to \qty{2.3}{eV} (\figref[f]{irradiation}). Note that the very high ion fluences of about \qty{5e7}{ions/spot} used in the study resulted in substantial milling of the hBN over an approximately 100~nm wide circular patch. It is also worth mentioning three interesting non-standard irradiation studies. In Ref.~\onlinecite{Glushkov2022}, optically active defects ensembles were created in hBN by combining a focused Xe-ion beam and subsequent DI-water etching. By using super-resolution optical microscopy combined with atomic force microscopy, the emitters were observed to be located at the rims of the etched patterns. Their broad spectra are centered at \qty{830}{nm}, matching emission spectra typically associated with {\VB} defects in other studies. In the second study, Ref.~\onlinecite{Ren2024}, evidence of carbon-based luminescent defects (emitting at \qty{2.15}{eV}) was reported after heterostructures comprising few-layer hBN covered with graphene on ${\rm SiO}_2$ were irradiated from the top by \qty{1}{keV} argon ions. Molecular dynamics simulations suggest that displaced carbon atoms could effectively be implanted into hBN, enhancing the likelihood of emission by carbon-based defects in hBN. In the third interesting non-standard irradiation study, Ref.~\onlinecite{Ren2025a}, the aim was again to create carbon-related color centers in a hBN flake, this time by first carbon irradiation (\qty{30}{keV} energy, \qty{10e14}{ions/cm^{-2}} fluence, and \qty{30}{\degree} incident angle) and then thermal annealing at \qty{1297}{K} in argon atmosphere for one hour, but they report only creation of {\VB} vacancies, thus far without clear signatures of single-photon emission.

\section{Ion Implantation}\label{sec:implantation}

\begin{figure*}[thb]
  \centering
  \includegraphics[width=2\columnwidth]{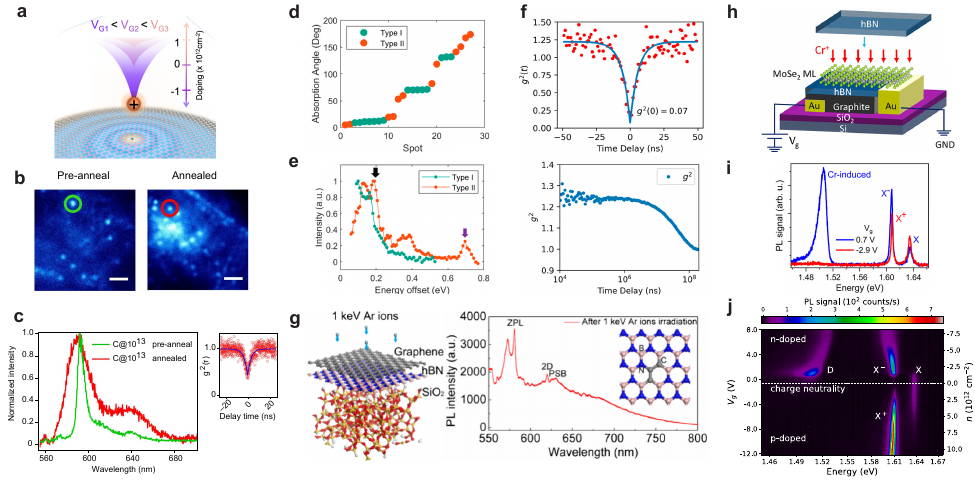}
  \caption{\textbf{Creation of SPEs utilizing ion implantation.} \textbf{a} A positively charged ion is accelerated towards a target 2D lattice. Reproduced with permission from \citet{Telychko2022}, Nano Lett. \textbf{22} (21), 8422–8429 (2022). Copyright 2022 American Chemical Society. \textbf{b} Integrated photoluminescence maps of carbon-implanted hBN films comparing pre-annealing and post-annealing conditions, with highlighted isolated SPEs. \textbf{c} PL spectra from the highlighted SPEs with an example $\mathrm{g^2}$ measurement. Reproduced with permission from \citet{Mendelson2021}, Nat. Mater. \textbf{20}, 321–328 (2021). Copyright 2020 Authors, under exclusive license to Springer Nature Limited. \textbf{d} PL absorption polarization characteristics and \textbf{e} photoluminescence excitation (PLE) spectra of $\mathrm{^{12}C^+}$ implantation-generated emitters in hBN (divided in two groups). \textbf{f} Second-order autocorrelation ($\mathrm{g^2}$) on a (upper) short and (lower) long time scale, highlighting the antibunching at $t=0$. The experimental data are normalized by the coincidence rate at longer timescales. Reproduced from \citet{Zhong2024}, Nano Lett. \textbf{24} (4), 1106–1113 (2024). Copyright 2024 Authors, licensed under a Creative Commons Attribution 4.0 (CC BY) License. \textbf{g} (left) Graphical representation of carbon implantation inside the hBN achieved via low-energy (\qty{1}{keV}) argon-ion beam irradiation of a graphene/hBN stack. (right) PL spectra of carbon-related emitter in the graphene/hBN stack, with highlighted the proposed zero-phonon line (ZPL) and phonon-side band (PSB), and the 2D Raman mode of graphene. The inset is a scheme of the proposed defect molecular structure ($\mathrm{C_BC_{N2}}$). Reproduced with permission from \citet{Ren2024}, ACS Appl. Nano Mater. \textbf{7} (3), 3436–3444 (2024). Copyright 2024, American Chemical Society. \textbf{h} Sketch of the device fabricated in Ref. \onlinecite{Bui2023} (Cr-implanted $\mathrm{MoSe_2}$ monolayer encapsulated in hBN and with a graphite backgate). \textbf{i} PL of the implanted device (at \qty{10}{K}) as a function of the backgate voltage. \textbf{j} PL spectra as a function of the backgate voltage with marked the doping type and calculated carrier concentration $n$. Reproduced from \citet{Bui2023}, ACS Appl. Mater. Interfaces \textbf{15} (29), 35321–35331 (2023). Copyright 2023 Authors, licensed under a Creative Commons Attribution 4.0 (CC BY) License.}
  \label{fig:implantation}
\end{figure*}

Color centers or optically active impurities can be easily introduced in 2D materials during the growth process \cite{Loh2024, Rivera2021}. For example in Ref.~\onlinecite{Liu2022}, carbon-related defects in monolayer hBN were created by chemical vapor deposition (CVD) synthesis at high-temperature. However, towards the integration of quantum emitters into photonic structures, ion implantation can provide a more controllable approach for fabricating type-specific and site-specific color centers. Ion implantation is normally referred to as the process of directly introducing impurities (for example, carbon, nitrogen, boron, silicon, or other dopants) into a material by irradiation with the respective ion species ~\cite{Bangert2013_NL, Susi2017, Tripathi2018, Bui2022, Villarreal2024, Willke2015, Telychko2022, Bangert2017}. Alternatively, impurities can be implanted indirectly upon irradiation with energetic particles by capturing displaced atoms into created vacancies ~\cite{Wang2012nl, Ma2014acsn}. For direct implantation of 2D materials, the ion energy should be relatively low (from a few tens to hundreds of eVs) to enable stopping and incorporation of the ions into the atomically thin target (see \secref{fundamentals}). For graphene (\figref[a]{implantation}), doping via ion implantation was extensively studied ~\cite{Willke2015, Susi2017, Tripathi2018, Villarreal2024}, and the successful implantation in monolayers was verified by atomic resolution high-angle dark field imaging (HAADF) with single-atom electron energy loss spectroscopy (EELS)~\cite{Bangert2013_NL} in STEM.

For implantation into hBN, so far the most attractive element has been -- mostly without competition -- carbon, because carbon is believed to be at the origin of many of the defect complexes studied~\cite{Mendelson2021, Gao2023, Zhong2024, Badrtdinov2023, Scholten2023, Singla2024, Stern2022, Stern2024, Guo2023, Chejanovsky2021, Ren2024, Fischer2023, Ren2025, Tawfik2017, Sajid2018, Mackoit-Sinkeviciene2019, Horder2025, Jara2021, Cholsuk2022, Cholsuk2024, Liu2022, Chatterjee2025}. For a more detailed discussion of possible candidates for the molecular structure of carbon-based defects, see \secref{nature}. Carbon-related luminescent defects are observed in hBN monolayers~\cite{Liu2022} and even in hBN nanotubes~\cite{Gao2023}. Moreover, site-specific electron microscopy imaging and EELS give further hints that carbon complexes are involved in photoluminescence (PL) emission~\cite{Singla2024}. In a pioneering study, \citet{Mendelson2021} achieved injection of carbon inside hBN via both carbon-incorporation during metal-organic vapour-phase epitaxy (MOVPE) growth and via irradiation with carbon ions at \qty{10}{keV} (ion fluence $10^{13}$ ions/cm$^2$). Notably, for defects grown by carbon-doped MOVPE, they showed both photon antibunching and magnetic activity (positive ODMR contrast up to 0.06\% with small zero-field spitting $D$). Concerning implantation, they irradiated both pristine MOVPE and exfoliated hBN with carbon (at \qty{10}{keV} energy and $10^{13}$ ions/cm$^{2}$ fluence). After implantation, antibunched quantum emitters with narrowed ZPL as compared to carbon-doped MOVPE were obtained (see \figref[b-c]{implantation}), and their density scaled with the ion fluence. Control irradiation with silicon and oxygen created no SPEs but rather only boron vacancies. Interestingly, subsequent annealing at \qty{1000}{\degreeCelsius} broadened the PL peaks (see \figref[b-c]{implantation}) and suppressed the single-photon behavior indicating the formation of defect ensembles. 

By contrast, \citet{Zhong2024} combined $\mathrm{^{12}C^+}$ irradiation (\qty{10}{keV} acceleration, \qty{0}{\degree} tilt angle, $10^{13}$ ions/cm$^{2}$ fluence under high vacuum) on commercial, exfoliated hBN with annealing in Ar gas at \qty{900}{\degreeCelsius} for 30 minutes. They report the generation of SPEs with a strong correlation between optical dipole orientation and crystal direction. In particular, for one of the two types of emitters reported, the absorption dipole orientation takes only values \qty{60}{\degree} apart (see \figref[d]{implantation}). The emitters exhibited highly repeatable excitation resonances in PLE and photon antibunching (see \figref[e,f]{implantation}). Further control irradiation with $\mathrm{^{16}O^+}$ confirmed that the reported emitters are not intrinsic defects, but rather induced by the carbon implantation process.
 
As an alternative approach, \citet{Ren2024} present carbon-related color centers (with ZPL around \qty{580}{nm}) generated in a graphene/monolayer hBN stack using \qty{1}{keV} argon-ion implantation without subsequent annealing (see \figref[g]{implantation} for a representative sketch). The created emitters overlap partly with Raman signatures from the graphene layer, but they generally show spectral features similar to the ones reported by \citet{Mendelson2021} (see \figref[c]{implantation}) with long photostability, although no antibunching measurements were reported. The idea of carbon doping in hBN by bare momentum transfer could provide a reliable and site-selective way to generate position-controlled, carbon-based emitters in hBN, and constitute an alternative to carbon doping via annealing~\cite{Koperski2020, Fischer2021, Badrtdinov2023, Karanikolas2023} or to heavy ion implantation~\cite{Lopez-Morales2021}. Future steps might, for example, involve combining the two approaches from \citet{Mendelson2021} and \citet{Ren2024}, by using a thicker layer of carbon and/or employing carbon irradiation at the same time. In Ref.~\onlinecite{Hua2024}, an amorphous, \qty{100}{nm} thick carbon mask was placed in front of freestanding hBN flakes (with thicknesses between 10 and 100~nm) to decelerate incoming carbon ions (\qty{40}{keV} acceleration voltage, \qty{2.5e14}{ions/cm^2}). This approach achieved the creation of two classes of emitters distributed around \qty{550}{nm} and \qty{590}{nm}, showing high stability, a short lifetime around \qty{1}{ns}, in-plane polarization in excitation, PL saturation, and photon antibunching. Currently, there are no experimental or simulated results on low-energy ion implantation into hBN. To this end, new works could be directed toward the optimization of the generation parameters. Remarkably, analytical potential molecular dynamics (APMD) simulations show that optimal parameters for carbon implantation into monolayer hBN are 50 eV ion energy combined with incident angles $\theta$ ranging from \qty{20}{\degree} to \qty{40}{\degree} \cite{Ren2025}. Further works seem to be needed to optimize the irradiation parameters for efficient implantation of 2D and related materials with low-energy ion implantation.

We note that a very recent and pioneering work successfully created magnetically active quantum emitters in the \qty{2}{eV} region by implantation. Specifically, they employed low-energy (\qty{2.5}{keV}) implantation of the magnetically active carbon isotope $^{13}$CO$_2$ into hBN\cite{Gao2025}. ODMR measurements showed that two types of defects were formed with spin $S=1/2$ and spin $S=1$ states, and spin coherence times in the range of tens to hundreds µs. They use hyperfine magnetic coupling between electronic and nuclear spin in $^{13}$CO$_2$ to fingerprint the presence of carbon in the created emitters. This represents an interesting advance for the determination of the microscopic origins of defect-related quantum emitters with magnetic response in 2D materials.

For TMDCs, very few works have utilized ion implantation for the creation of optically active defects so far. Direct implantation into monolayer TMDCs is particularly challenging because of their increased structural sensitivity resulting in instability under implantation conditions and the tendency of high energy ions to penetrate and stop inside the substrate. Therefore, efficient ion-implantation into monolayers requires typically much lower ion energies than in conventional implantation processes (see \secref{fundamentals}). Nevertheless, \citet{Bui2023} successfully demonstrated implantation of low-energy $\mathrm{^{52}Cr^+}$ ions (\qty{25}{eV}, \qty{220}{\degreeCelsius}) directly into a monolayer MoSe\textsubscript{2} field-effect structure (\figref[h]{implantation}). The low-energy implantation resulted in a well-defined sub-gap PL (\figref[i]{implantation}), although no single-photon emission was observed. Since the defect PL was only present for the $n$-doped regime (\figref[j]{implantation}), the radiative recombination of a localized electron in a deep acceptor state with a free valence band hole was proposed (cf. \secref{defects}) as the potential origin. An alternative route, in particular to create quantum emitters in the telecom range with long coherence times, may be the implantation of rare earth dopants, such as Er\cite{Gritsch2022}, where the photophysics are dictated by internal transitions of the partially filled 4f-shell and are only weakly modified by the host crystal. In this context, a recent study reported evidence for narrowband PL emission in the telecom range with very long lifetime (\qty{4.5}{ms} at room temperature) after implantation of Er dopants into thick (above 200 nm) WS\textsubscript{2} flakes followed by annealing at \qty{400}{\degreeCelsius}\cite{Garcia-Arellano2025}.

\section{E-beam irradiation}
\label{sec:e-beam}

\begin{figure*}[thb]
  \centering
  \includegraphics[width=1.95\columnwidth]{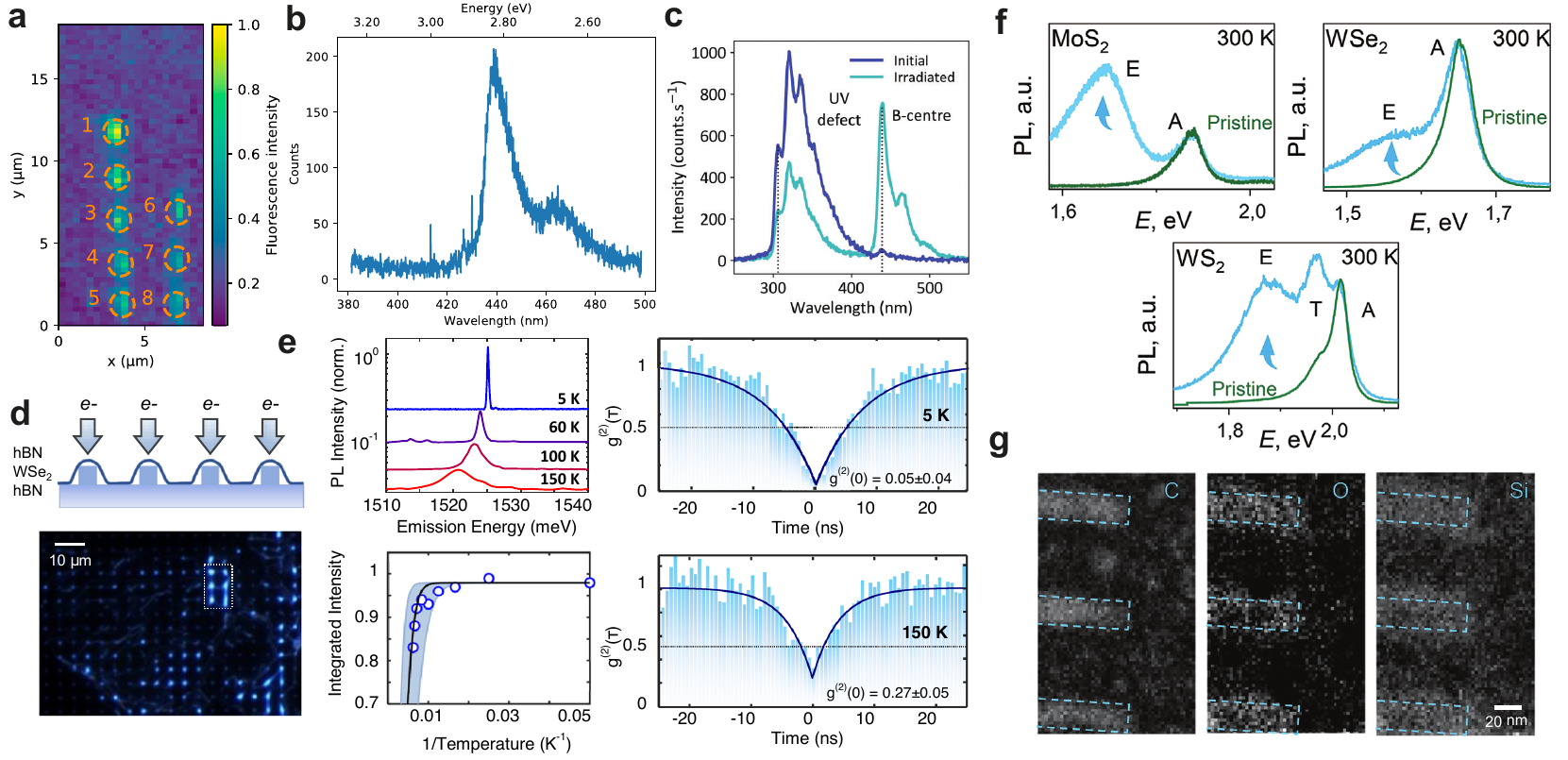}
  \caption{\textbf{Single-photon emission in hBN and TMDCs generated by electron beam irradiation.} \textbf{a} PL map of a SPE array generated via extended e-beam exposure (15~keV) of hBN. \textbf{b} Associated PL spectrum of a given spot from the array in \textbf{a} exhibiting the ZPL and PSB of a blue emitter. Reproduced with permission from \citet{Fournier2021}, Nat. Commun. \textbf{12}, 3779 (2021). Copyright 2021 Authors, licensed under a Creative Commons Attribution (CC BY 4.0) License. \textbf{c} CL spectra of UV and blue emission before and after the \qty{300}{s} e-beam irradiation. The UV emission decreases while the blue emission increases. Reproduced with permission from \citet{Nedic2024}, Adv. Optical Mater. \textbf{12}, 2400908 (2024). Copyright 2021 Authors, licensed under a Creative Commons (CC BY-NC-ND) License. \textbf{d} Combined e-beam and strain activation of telecom-range SPE in WSe\textsubscript{2}. \textbf{e} PL spectrum of a spot from the array in \textbf{d} for temperatures between 5 K and 150 K. The PL peak quenches at 150K. Associated second-order correlation functions at 5 K and 150 K. Adapted with permission from \citet{Parto2021}, Nat. Commun. \textbf{12}, 3585 (2021). Copyright 2024 Authors, licensed under a Creative Commons (CC BY) License. \textbf{f} PL spectra for multiple TMDCs showing a consistently E peak after e-beam irradiation. \textbf{g} EELS maps for C, O, and Si for MoS\textsubscript{2} sample in \textbf{f}. The e-beam patterned arrays, highlighted with blue-dashed lines exhibits how the C, O, and Si content is larger within these regions. Adapted with permission from \citet{Yagodkin2022}, Adv. Funct. Mater. \textbf{32}, 2203060 (2022). Copyright 2022 The Author(s), licensed under a Creative Commons Attribution Non Commercial 4.0 (CC BY-NC) License.}
  \label{fig:e-beam}
\end{figure*}

As an alternative defect generation technique, focused electron irradiation has attracted widespread attention. The method makes use of the focused electron beam in standard scanning (transmission) electron microscopes where typical electron energies range from \qty{1}{keV} to more than \qty{100}{keV}. Depending on their kinetic energy, electrons introduce damage to the target material via knock-on damage, inelastic scattering, and beam-induced chemical etching (cf. \secref{fundamentals}). Theoretical studies indicate that for $\mathrm{MoS}_2$ and hBN\cite{Susi2019_NatRevPhys} radiolysis and chemical etching are the primary factors causing damage at electron beam energies below 70 keV. In the latter case, electrons generate broadband electromagnetic fields that ionize the surrounding matter. Knock-on damage dominates at beam energies exceeding 90 keV for $\mathrm{MoS}_2$ and 115 keV for hBN~\cite{Cretu2014}. As noted in \secref{fundamentals}, emitters in both $\mathrm{MoS}_2$ and hBN have been experimentally created or activated below these theoretically estimated thresholds.

Prolonged focused electron beam irradiation at an energy of \qty{15}{keV}, which is significantly below the knock-on damage threshold \cite{Bui2023a}, was shown to result in site-selective, room temperature single photon emission (\figref[a]{e-beam}). These sites emit photons with a characteristic blue emission spectrum centered at \qty{440}{nm} (\figref[b]{e-beam}) \cite{Fournier2021}. An earlier cathodoluminescence study reported similar blue emitters \cite{Shevitski2019}, which could be activated or deactivated under the electron beam irradiation and that appeared simultaneously with single-photon emission centers in UV \cite{Bourrellier2016}. More recent cathodoluminescence experiments \cite{Gale2022} propose that blue emitters derive from UV-active defects which are modified under the electron beam due to a bond-breaking–restructuring process of carbon-related defect complexes (\figref[c]{e-beam}). A more recent work from the same group suggests that the electron beam can generate electromigration of charged defects leading to clustering and formation of B-centers~\cite{Nedic2024} (see also \secref{nature}). Interestingly, electron beam irradiation (\qty{3}{keV}) was also reported to result in the formation of position-controlled SPEs at \qty{575}{nm} \cite{Kumar2022a}, but to date the results have not been reproduced independently.

In the context of single-photon emission in TMDCs, electron beam irradiation has so far been mostly used to introduce active defects sites in strain-patterned WSe\textsubscript{2} \cite{Parto2021, Xu2022, Paralikis2024, Piccinini2024}. \figref[d]{e-beam} shows an example of a strain-patterned array of single-photon emitters in monolayer WSe\textsubscript{2} with emission lines around \qty{1.55}{eV} (\figref[e]{e-beam}), whereby the strain sites were activated by focused electron beam irradiation (\qty{100}{keV}). Notably, the latter study demonstrated single-photon statistics up to a temperature of \qty{150}{K}. The rationale behind the electron beam activation process is to introduce locally in-gap defect states that serve as recombination centers for strain-localized (dark) exciton states (cf. \secref{defects} and \secref{nature}). Although single-photon emitters in WSe\textsubscript{2} were initially reported without any additional electron beam exposure \cite{Koperski2015, Srivastava2015,He2015,Tonndorf2015, Palacios-Berraquero2016,Branny2016}, several recent studies emphasize the benefit of using electron beam exposure to further optimize the emitter creation process\cite{Parto2021, Xu2022, Paralikis2024, Piccinini2024}. Different defect creation mechanisms may prevail depending on the used electron energy (cf. \secref{fundamentals}). Above the knock-on damage threshold \cite{Parto2021, Paralikis2024, Piccinini2024}, creation of intrinsic defects, such as chalcogen vacancies, may be dominant, while below the knock-on damage threshold\cite{Moody2018, Yagodkin2022,Xu2022} chemical etching and electron beam-induced deposition may create both intrinsic and extrinsic defects, such as adsorbed or chemisorbed carbonaceous species, in the TMDC layer. In this context, Ref. \onlinecite{Yagodkin2022} revealed that electron beam exposure can alter the photoluminescence emission characteristics of MoS\textsubscript{2}, WSe\textsubscript{2}, and WS\textsubscript{2}. After exposure, all materials exhibited a broad defect luminescence at energies below the band-gap related exciton luminescence (\figref[f]{e-beam}), notably at room temperature and without the demonstration of single-photon emission. For MoS\textsubscript{2}, compositional analysis by EELS (\figref[g]{e-beam}) further revealed a uniform Mo and S distribution, while significant C, O, and Si contents were detectable in the electron beam exposed regions \cite{Yagodkin2022}. After mechanical cleaning by AFM\cite{Rosenberger2018}, the room-temperature defect luminescence disappeared. Both observations are consistent with an extrinsic contribution to the defect luminescence due to e-beam activated deposition of organic species, which can be omitted by encapsulating the TMDCs e.g. in hBN before the beam exposure takes place \cite{Yagodkin2022, Klein2021a}. For monolayer MoS\textsubscript{2}, electron beam irradiation with \qty{5}{keV}, i.e. far below the knock-on threshold, was recently shown to result in the formation of single photon emitters\cite{Dash2025}, whose spectral characteristics are consistent with vacancy-related SPEs created by 30 keV He ion beam irradiation\cite{Klein2021a,Mitterreiter2020,Mitterreiter2021}. Importantly, the electron beam irradiation was found to be only effective on non-encapsulated samples suggesting beam-induced chemical etching or chemical modification of the exposed MoS\textsubscript{2} layer to be the dominant process of defect creation. The above discussion also highlights the need to carefully distinguish between creation of defect emitters, e.g. by the introduction of lattice defects, or the activation of defect emitters, e.g. by modifying the charge state of already existing defects.

\section{X-Ray and UV-Irradiation}\label{sec:xray}

\begin{figure}[thb]
  \centering
  \includegraphics[width=.975\columnwidth]{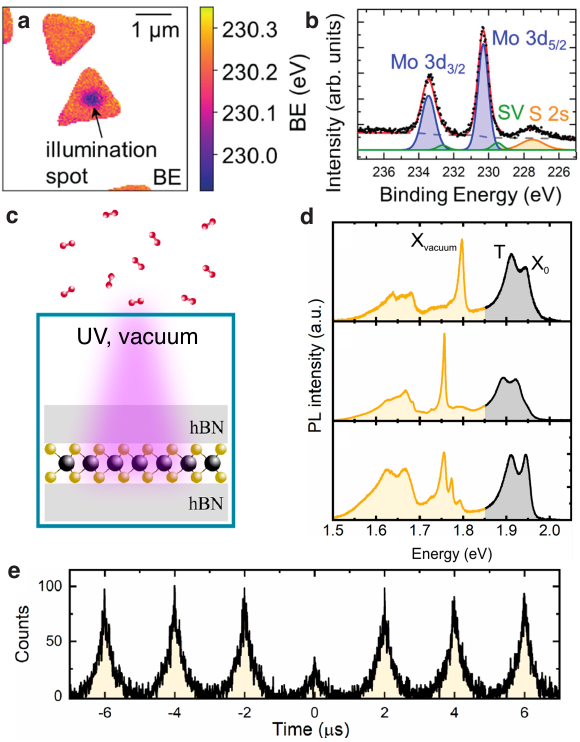}
  \caption{\textbf{Local generation of defects utilizing X-ray irradiation.} \textbf{a} Map of the binding energy of the Mo 3d\textsubscript{5/2} core-level photoemission peak after the illumination with a focused X-ray beam, indicating the local generation of defects. \textbf{b} Core-level photoemission spectra of the spin-orbit split doublets Mo 3d\textsubscript{5/2}/Mo 3d\textsubscript{3/2}  for an X-ray dose of \qty{5}{MJ/cm^2}. The extracted contributions of the X-ray generated single-sulfur vacancies (SV) and disulfur vacancies (S 2s) are highlighted in green and yellow, respectively. Reprinted with permission from~\citet{Grunleitner2022}, ACS Nano \textbf{16}, 20364–20375 (2022). Copyright 2022 American Chemical Society. \textbf{c} Schematic of defect generation within a encapsulated monolayer under UV illumination in vacuum. \textbf{d} Representative spectra of defect emitters (yellow) created via UV illumination. \textbf{e} The defect emitters show single photon characteristics. Reprinted with permission from~\citet{Wang2022}, ACS Nano \textbf{16}, 21240–21247 (2022). Copyright 2022 American Chemical Society.}
  \label{fig:xray}
\end{figure}

So far, only very few attempts have been reported to locally generate defects and corresponding SPEs in TMDCs and hBN with the help of high energy photon irradiation. Recently, the X-ray based generation of sulfur vacancies was reported for monolayer MoS\textsubscript{2} (\figref[a]{xray}) ~\cite{Grunleitner2022}, which are consistent with the SPEs in the same material\cite{Mitterreiter2020, Mitterreiter2021}. Soft X-ray irradiation leads to the emergence of distinct 3d Mo spectral features in the core-level photoemission spectra of MoS\textsubscript{2} associated with undercoordinated Mo atoms (\figref[b]{xray}). In particular, the photoemission spectra of the spin-orbit split doublet Mo 3d\textsubscript{5/2} and Mo 3d\textsubscript{3/2} (blue fits in \figref[b]{xray}) and the corresponding appearance of an additional doublet shifted to lower binding energies (green) can be traced to the impact of the X-ray irradiation and to the presence of sulfur monovacancies (SV). Generally, similar spectral changes have been observed earlier for Ar and He bombardment~\cite{McIntyre1990, Huang2019}, thermal annealing~\cite{Donarelli2013, Zhang2020}, and as-grown defective TMDCs~\cite{Kastl2019}. Monitoring the binding energy of the Mo 3d\textsubscript{5/2} core-level photoemission peak across the MoS\textsubscript{2} monolayers indicates the emergence of lateral space-charge regions associated with the local, X-ray based generation of sulfur vacancies in MoS\textsubscript{2} (\figref[a]{xray}). Most likely, the sulfur monovacancies are locally generated via ionization-induced bond destabilization arising from the X-ray absorption within MoS\textsubscript{2} as well as inelastic scattering of secondary electrons emitted from the substrate~\cite{Grunleitner2022}. The latter secondary process is also relevant for the generation of defects in TMDCs by charged particle beams (cf. corresponding sections above). 

We further highlight a recent study demonstrating that SPEs can be induced in monolayer MoS\textsubscript{2} by unfocused UV irradiation with a deuterium lamp over an extended period of time in vacuum (\figref[c]{xray}).~\cite{Wang2022} Interestingly, the resulting spectral signatures (defect peak centered around \qty{1.75}{eV}, \figref[d]{xray}) are consistent with those reported for He-ion beam-irradiated samples (see \secref{ion}) as well as thermally treated samples that are associated with the generation of pristine, unpassivated sulfur vacancies~\cite{Klein2019c, Klein2021a, Barthelmi2020, Hoetger2023, Mitterreiter2021}. For defects created by UV irradiation of encapsulated samples in vacuum, a clear antibunching and corresponding single photon emission was demonstrated (\figref[e]{xray}). Another recent study reported creation of defect emission in multilayer GaSe by UV laser (266 nm) irradiation, although without evidence for single photon emission \cite{Varghese2025}.

\section{Microscopic nature of the emitters}
\label{sec:nature}

\begin{figure*}[thb]
  \centering
  \includegraphics[width=1.95\columnwidth]{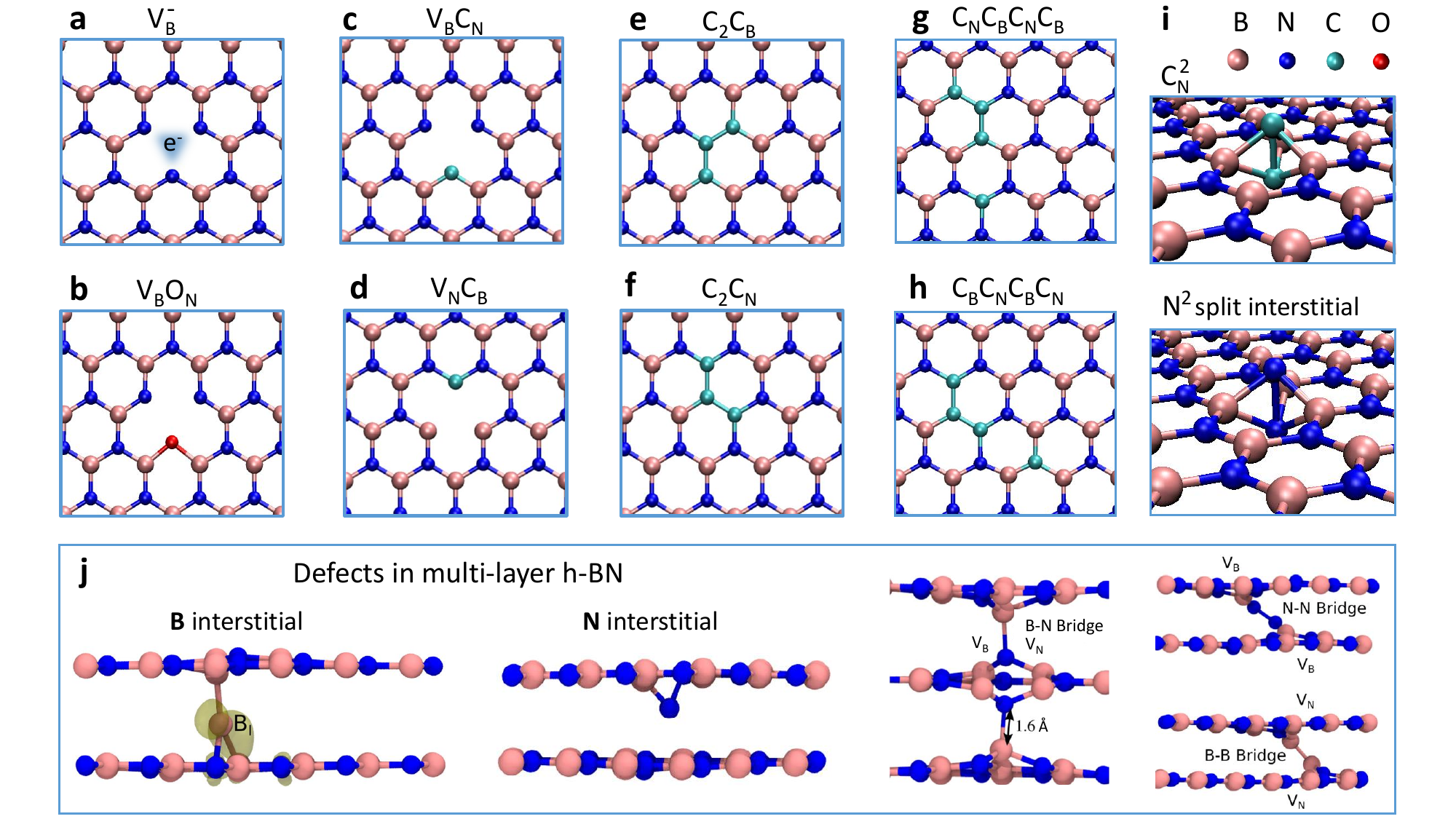}
  \caption{\textbf{Microscopic structure of several candidate defects for experimentally observed SPEs in mono- and multilayer hBN.} \textbf{a} Charged boron vacancy. \textbf{b} Oxygen-vacancy center. \textbf{c-h} Examples of carbon-based centers. \textbf{i} Examples of split interstitial centers. \textbf{j} Examples of interlayer defect centers in multi-layer hBN. The atomic models of the defect structures in \textbf{j} are taken from Ref.~\onlinecite{Strand2020}.} 
  \label{fig:microscopic}
\end{figure*}
\emph{For hBN}. As mentioned in previous reviews~\cite{Aharonovich2022}, the quantum emitters in hBN are typically categorized into four main families: the broad emission in the NIR, the sharper emission around \qty{2}{eV} displaying identifiable phonon sidebands, the blue emitters with emission around \qty{2.8}{eV} (the so-called B-centers), and finally the emission in the UV around \qty{4.1}{eV}. There is strong evidence through photoluminescence excitation spectroscopy and \emph{ab-initio} calculations that these four families of emitters are localized color centers with energies of their electronic states deep in the large optical gap (\qty{5.9}{eV}) of hBN. This situation corresponds to the atomic picture presented in \figref[a]{defect_models}. Efficient excitation of these deep electronic states is possible by exciting on resonance discrete vibronic states, i.e. electronic states dressed by the vibrations surrounding the emitter, as revealed by PLE measurements on NIR \cite{Baber2022}, 2eV~\cite{Grosso2020, Fischer2023}, and B-centers~\cite{Horder2025}. In this section, we focus on the emitters created by irradiation engineering and we attempt to relate the mechanism of creation to their proposed microscopic structure according to the published literature. Understanding the effect of particle irradiation on the atomic structure of hBN can shed light on the internal structure of emitters, a topic for which to date no unified consensus has been reached. \tabref{hbn} shows a survey of the main irradiation methods used and links them to the photophysical properties of the created emitters.

\begin{table*}[tphb]
  \centering
  \begin{tabular}{>{\centering\arraybackslash}p{1.65cm} >{\centering\arraybackslash}p{2.5cm} >{\centering\arraybackslash}p{1.5cm} >{\centering\arraybackslash}p{2.5cm} >{\centering\arraybackslash}p{2cm} >{\centering\arraybackslash}p{2cm} >{\centering\arraybackslash}p{2cm} >{\centering\arraybackslash}p{2.2cm}}
  \toprule
    
    Irradiation Method & \multirow{2}{*}{Particle} & \multirow{2}{*}{Ref.} & \multirow{2}{*}{$\mathrm{E_{ZPL}}$ [eV]} & \multirow{2}{*}{FWHM [GHz]} & \multirow{2}{*}{Lifetime [ns]} & \multirow{2}{*}{$\mathrm{g^{(2)}(0)}$} & \multirow{2}{*}{Proposed origin} \\  
     \cmidrule(r){1-8}
     & He & \cite{Gao2021} & $\approx1.5$ & N.A. & N.A. & N.A. & $\mathrm{V_B^-}$ \\
     \cmidrule(r){2-8}
     \multirow{5}{*}{\shortstack{Plasma and\\ low-energy\\ ion}} & O & \cite{Vogl2019} & $\approx1.91-2.25$ & 2100-3100 & 0.55-0.8 & 0.14-0.23 & N.A. \\
     \cmidrule(r){2-8}
     & \multirow{2}{*}{O} & \multirow{2}{*}{\cite{Fischer2021,Fischer2023}} & \multirow{2}{*}{$\approx1.8-2.3$} & \multirow{2}{*}{500 (\qty{10}{K})} & \multirow{2}{*}{2} & \multirow{2}{*}{0.1-0.8} & $\mathrm{V_N C_B}$, $\mathrm{C_2 C_N}$, $\mathrm{C_2 C_B}$ \\
     \cmidrule(r){2-8}
     & Ar, O & \cite{Xu2018} & $\approx1.6-2.25$ & 1660 (\qty{10}{K}) & 2.4 & 0.1 & $\mathrm{V_B O_2}$ \\
     \cmidrule(r){1-8}
     & Li, Ga, neutr. & \cite{Gottscholl2020} & $\approx1.46$ & N.A. & 1.2 & N.A. & $\mathrm{V_B^-}$ \\
     \cmidrule(r){2-8}
     \multirow{2}{*}{\shortstack{Ion\\ irradiation}} & He, O, N, Xe, Ar& \cite{Kianinia2020, Liang2023, Zabelotsky2023,Carbone2025} & $\approx1.5-1.55$ & 38k (\qty{5}{K})\cite{Kianinia2020} & 0.77-1.028\cite{Carbone2025} & N.A. & $\mathrm{V_B^-}$ \\
     \cmidrule(r){2-8}
     & Ga & \cite{Ziegler2019, Klaiss2022} & $\approx2-2.3$ & N.A. & N.A. & 0.24, 0.33 & $\mathrm{V_B O_N^-}$\cite{Li2022b} \\
     \cmidrule(r){2-8}
     & He & \cite{Liang2023c} & 2.73, 2.86 & N.A. & N.A. & N.A. & $\mathrm{B_i}$ \\
     \cmidrule(r){1-8}
     & electrons & \cite{Murzakhanov2021} & $\approx1.55$ & N.A. & N.A. & N.A. & $\mathrm{V_B^-}$ \\
     \cmidrule(r){2-8}
     & - & \cite{Tran2016a} & $\approx1.6-2.2$ & 283-936 & 2.9-6.7 & 0.34-0.39 & $\mathrm{N_B V_N}$ \\
     \cmidrule(r){2-8}
     \multirow{3}{*}{E-beam} & \multirow{2}{*}{-} & \multirow{2}{*}{\cite{Kumar2022a}} & \multirow{2}{*}{$\approx2.15$} & \multirow{2}{*}{$\approx18$k} & \multirow{2}{*}{3-5} & \multirow{2}{*}{0.2-0.55} & $\mathrm{(C_B C_N)^2}$ $\mathrm{(C_N C_B)^2}$ \\
     \cmidrule(r){2-8}
     & - & \cite{Singla2024} & $2.16$ & 6800 & N.A. & N.A. & Carbon cluster \\
     \cmidrule(r){2-8}
     & - & \cite{Fournier2021,Fournier2023a,Gerard2025} & $\approx2.85$ & 0.75 (5K)\cite{Gerard2025} & 1.85 & 0.1-0.25 & $\mathrm{C^2_N}$, $\mathrm{N^2}$, \emph{trans}-$\mathrm{C}_{4}$ \\
     \cmidrule(r){2-8}
     & - & \cite{Gale2022,Horder2022} & $\approx2.85$ & $\approx2.2$ (5K)\cite{Horder2022} & 2.27\cite{Horder2022} & 0.19 & $\mathrm{C^2_N}$, $\mathrm{N^2}$, \emph{trans}-$\mathrm{C}_{4}$ \\
     \cmidrule(r){1-8}
     & C & \cite{Mendelson2021} & $\approx2$ & 4424 & 2-6 & <0.5 & $\mathrm{V_B C_N^-}$ \\
     \cmidrule(r){2-8}
     \multirow{3}{*}{Implantation} & Ar on Graph./hBN & \cite{Ren2024} & 2.14 & N.A. & N.A. & N.A. & $\mathrm{C_B C_{N2}}$ \\
     \cmidrule(r){2-8}
     & C & \cite{Zhong2024} & 2.24 & <36 (4 K) & 4.4 & 0.07 & DAP \\
     \cmidrule(r){2-8}
     & \multirow{2}{*}{$^{13}$CO$_2$} & \multirow{2}{*}{\cite{Gao2025}} & \multirow{2}{*}{2.1 and 1.8} & \multirow{2}{*}{$\approx$25k} & \multirow{2}{*}{N.A.} & \multirow{2}{*}{<0.5} & $C^+_BC^0_N$-DAP $C_BO_N$ \\
     
    \bottomrule
  \end{tabular}
  \caption{Table presenting photophysical information about quantum emitters created by irradiation engineering in hBN. For each category of emitters, the proposed defect complex structure is listed. N.A.: not available. DAP: donor-acceptor pair.}
  \label{tab:hbn}
\end{table*} 

Generally speaking, identifying the microscopic nature of quantum emitters in multilayer hBN is challenging. Atomic resolution imaging techniques, such as transmission electron microscopy (TEM) and scanning tunneling microscopy (STM), often encounter non-ideal experimental conditions with hBN samples. Despite promising early attempts using TEM and cathodoluminescence~\cite{Hayee2020}, TEM imaging faces difficulties due to the bulkiness of samples and the low atomic number $\mathrm{Z}$ of boron and nitrogen elements. The latter renders identifying atomic defects within atomic columns difficult, since the intensity of the detected signal is proportional to $\mathrm{Z}^{1.8-1.9}$ resulting in poor contrast for single point defects in a multilayer sample. Additionally, correlating type and structure of defects with their photophysical properties simultaneously is problematic. Often, optical spectroscopy is used to detect quantum emitters, followed by TEM for atomic resolution characterization~\cite{Singla2024, Liang2023c}, but the difference in spatial resolution between the two techniques makes it uncertain whether  the same defects are predominantly probed in the different experiments. Sometimes different samples are used for optical and electron microscopy, further complicating direct correlation.
To create a tunneling current, STM requires few-layer or monolayer hBN samples due to its large band gap. Early studies using graphene-covered hBN~\cite{Wong2015} provided insights into defect charge states and electronic structures but, unlike a more recent work~\cite{Qiu2024}, did not correlate these with optical measurements. In \secref{atomistic} we highlight recent developments combining STM and luminescence at the atomic scale that may tremendously accelerate the identification of the microscopic nature of SPEs in vdW materials.

Nevertheless, by combining multiple, complementary spectroscopic methods, the nature of some emitters in hBN was already clarified. \figref{microscopic} sketches ball-and-stick models of defect structures occurring in hBN as discussed in the following. Combining photoluminescence and magnetic measurements, including ODMR and electron paramagnetic resonance spectroscopy (EPR), was used to pin-point the nature of the broadband emitters in the NIR with an emission peak centered around \qty{800}{nm} wavelength~\cite{Gottscholl2020}. Negatively charged boron vacancies $\mathrm{V_B^-}$ (\figref[a]{microscopic}) are now generally accepted to be responsible for this broadband NIR emission line. Furthermore, the broad luminescence spectrum featuring large phonon sidebands and the apparent absence of a zero-phonon line was related to the large electron-phonon coupling in {\VB} defects~\cite{Ivady2020a}. The comparably large Huang-Rhys factors above $3.5$ explains that most of the emission is going through the phonon sidebands~\cite{Libbi2022}. Ensembles of {\VB} are reliably generated via irradiation methods (see \secref{ion}). It should be pointed out that the concentration of optically active defects may be substantially lower than what is expected just after irradiation. For example, boron vacancies may be neutral or doubly-charged~\cite{Weston2018, Strand2020, Fraunie2025} due to charge transfer from other defects, or defect concentration can decrease due to the annealing of defects through annihilation of vacancies with interstitials, which are mobile at room temperature~\cite{Weston2018, Strand2020}. So far, {\VB} defects have been thoroughly characterized only as ensembles, since their low internal quantum efficiency hinders the detection and characterization of single defects~\cite{Clua-Provost2024}. Furthermore, the generation mechanism via irradiation engineering and the mechanism leading to the trapping of an electron by the boron vacancy is still under debate and will require further studies to be clarified~\cite{Carbone2025}. Nonetheless, the {\VB} defect is considered to be one of the few hBN defects whose internal atomic structure and level scheme are well known, thanks to the extensive amount of studies conducted to date~\cite{Gottscholl2020, Ivady2020a, Reimers2020, Baber2022, Mu2022, Gong2023, Udvarhelyi2023, Durand2023, Carbone2025}. The system, possessing $\mathrm{D_{3h}}$ point symmetry, exhibits a spin triplet ($\mathrm{S=1}$) ground state ${}^3A_2'$ and a corresponding spin-triplet excited state ${}^3E''$~\cite{Gottscholl2020, Clua-Provost2024}. The ground state presents a so-called axial zero-field splitting $\mathrm{D_g \sim}$ \qty{3.47}{GHz} between the spin sublevels $m_s=0$ and $m_s=\pm 1$, with $m_s$ electron spin projections along the $c$-axis of the hBN~\cite{Gottscholl2020}. Likewise, the excited-state spin triplet has a zero-field splitting $\mathrm{D_e \sim}$ \qty{2.1}{GHz}~\cite{Baber2022, Mathur2022, Yu2022}. Additionally, a manifold of metastable singlet states (MS) is present, which can be populated from the excited state via intersystem crossing. Spin transitions between different $m_s$ states are normally addressed by microwave pumping in ODMR experiments, where such transitions appear as strong resonances in ODMR spectra~\cite{Gottscholl2020}. It is generally agreed that the splitting between those resonances depends on magnetic fields through the Zeeman effect~\cite{Gottscholl2020}, whereas the residual splitting $E$ (in absence of external magnetic field) has recently been under debate. Originally, it was attributed to strain effects in the hBN~\cite{Gottscholl2020, Guo2022}, but recent works discuss contributions from spin-sublevel mixing induced by a local electric field perpendicular to the $c$-axis~\cite{Gong2023, Udvarhelyi2023, Durand2023}. Lastly, each ODMR resonance is split seven-fold due to hyperfine interaction with the three surrounding $^{14}\mathrm{N}\, I = 1$ nitrogen nuclei\cite{Gottscholl2020}. It is assumed that only the excited state of the triplet system is optically active, while non-radiative decay paths are spin dependent, that is, spins excited from the $m_s = \pm 1$ manifold decay radiatively less likely compared to the excited population of the ${m_s=0}$ state. Therefore, {\VB} defects can be spin-polarized in the $m_s=0$ spin sub-level within a few optical cycles~\cite{Gottscholl2020, Clua-Provost2023}. Thanks to the fairly short lifetime of the metastable state (on the order of tens of nanoseconds~\cite{Whitefield2023, Clua-Provost2024}) the defect relaxes rather rapidly and can be re-pumped efficiently, despite its poor quantum efficiency. The {\VB} can therefore be optically initialized and coherently manipulated~\cite{Gottscholl2020, Gottscholl2021}. Typical spin–lattice relaxation times $T_1$ are on the order of tens of µs~\cite{Gottscholl2021, Gao2021, Guo2022, Haykal2022, Ren2025a} and spin coherence times $T_2$ are around 100~ns~\cite{Gao2021, Gong2023, Rizzato2023, Haykal2022, Ren2025a}, with the possibility to be extended up to few µs~\cite{Rizzato2023}. Yet, the coherence time of {\VB} defects is limited by the nuclear spin bath of the hBN crystal, reducing their competitive edge against similar spin defects in 3D host systems ~\cite{Konrad2025}. A recent work using high temperature annealing in oxygen reported quantum emitters with emission around \SI{1.5}{eV} and very narrow line width~\cite{Mohajerani2024}. If these emitters are paramagnetic, they could represent a good contender for spin-state manipulations on the single photon and emitter scale.

Next, we focus our discussion towards the origin of the SPEs observed around \qty{2}{eV}. These emitters usually display an identifiable ZPL with clear phonon sidebands at energies around 160-170 meV below the ZPL. The phonon sidebands usually comprise both bulk optical phonon modes~\cite{Wigger2019} and localized phonon modes~\cite{Fischer2023}. Many types of elements and defect structures have been proposed as the microscopic origin of the ZPLs around \qty{2}{eV}. This includes silicon~\cite{Sajid2018}, oxygen (\figref[b]{microscopic})~\cite{Sajid2018,Li2022b}, and most prominently carbon (\figref[c-h]{microscopic})~\cite{Sajid2018, Li2022a,Jara2021,Golami2022,Sajid2020}, all of which are typically ubiquitous elements present around the hBN sample. A first systematic study involving the growth of hBN with MOVPE and MBE in carbon-rich environments as well as carbon implantation (cf. \figref[b]{implantation}) showed that carbon-based defects are the likely source for the emission observed around \qty{2}{eV}~\cite{Mendelson2021}. In Ref. \onlinecite{Sajid2018}, the authors compared four different carbon-based structures with different charge and spin states, namely $\mathrm{C_B}$, $\mathrm{C_N}$, $\mathrm{V_BC_N}$ (\figref[c]{microscopic}), and $\mathrm{V_NC_B}$ (\figref[d]{microscopic}). The $\mathrm{V_BC_N}$ was identified as the most likely explanation of the observed luminescence spectra ~\cite{Sajid2018}. A further study prepared comparable quantum emitters in hBN via low-energy oxygen irradiation and subsequent annealing~\cite{Fischer2021}. Ref.~\onlinecite{Fischer2021} explained the formation of these carbon-based defect complexes by a two-step process. First, a low-energy ion irradiation amorphizes the top few layers of hBN. Then, the subsequent annealing in a carbon-rich environment leads to the formation of defects with substitutional carbon atoms, such as $\mathrm{V_NC_B}$. An extension of this work, using the same irradiation-based method, compared low-temperature PL spectra with 26 optical transitions calculated from nine different defect complexes with different spin and charge states~\cite{Fischer2023}. Their model used a novel combination of \emph{ab-initio} calculations with open quantum system theory (OQST), in particular the polaron model~\cite{Nazir2016}, allowing an accurate description of the coupling of the electronic states with phonons in the presence of the electromagnetic field~\cite{Iles-Smith2017}. A statistical comparison of the theoretical spectral profiles for these 26 different optical transitions with the experimental data from 12 emitters in hBN suggested that $\mathrm{C_2C_B}$ (\figref[e]{microscopic}), $\mathrm{C_2C_N}$ (\figref[f]{microscopic}) and $\mathrm{V_NC_B}$ defect complexes match the experimental data best~\cite{Fischer2023}. The above theoretical approach further allows modeling of photoluminescence excitation spectroscopy (PLE) maps,~\cite{Fischer2023} which give crucial information about absorption properties and the electron-phonon coupling strength of the emitters.~\cite{Grosso2020} When comparing the calculated PLE maps with experimental data it was found that, in fact, $\mathrm{C_2C_N}$ and $\mathrm{C_2C_B}$ captured well the absorption spectra measured with PLE, with better agreement than $\mathrm{V_NC_B}$. These results were in line with a comparable work using emitters in hBN produced by high temperature annealing~\cite{Tieben2023}. We note that calculations of $\mathrm{C_2C_N}$ lifetimes agree well with measurements on \qty{2}{eV} emitters, showing nanosecond lifetimes (see \tabref{hbn}), while calculations on $\mathrm{C_2C_B}$ have shown two orders of magnitude longer lifetimes~\cite{Li2022a}. These works confined their comparison between experiments and theory to dimers and carbon trimers. More recently, four-atom structures were proposed as the microscopic origin of \qty{2}{eV} emissions generated by electron-beam irradiation~\cite{Kumar2022a}. Other even larger complexes, such as $\mathrm{C_NC_BC_NC_B}$ (\figref[g]{microscopic}) or $\mathrm{C_BC_NC_BC_N}$ (\figref[h]{microscopic}), were also proposed as the emitters generated by electron-beam irradiation~\cite{Singla2024}, annealing at high temperatures~\cite{Li2023}, or chemical vapor deposition~\cite{Tang2025}. More recently, a work employing pulsed laser deposition (PLD) combined with C-doping has proposed a trimer structure ($\mathrm{C_BC_NC_B}$) for the reported SPEs (showing values for autocorrelation among the best ones for SPEs in hBN \cite{Chatterjee2025}).
Thus it seems that many different carbon complexes can be formed in the hBN lattice, thereby making the identification and controlled generation of \qty{2}{eV} quantum emitters in hBN challenging. Further systematic studies are necessary and new methodologies need to be developed to identify the true microscopic nature of these emitters and to find ways to generate them on demand and site-selectively, ideally with high yield, specificity, and accuracy. Magnetically active quantum emitters in the \qty{2}{eV} range have been now observed in samples synthesized in carbon-rich MOVPE~\cite{Stern2019, Stern2022, Stern2024}. As addressed in \secref{implantation}, a recent work reports the successful creation of apparently spin-active quantum emitters in the \qty{2}{eV} region via low-energy implantation of $\mathrm{CO_2}$ purified with the $^{13}\mathrm{C}$ isotope \cite{Gao2025}. \emph{Ab-initio} calculations compared with magnetic spectroscopy suggest that the microscopic origin of these defects are the so-called $C^+_BC^0_N$ donor-acceptor pair (DAP) and $C_BO_N$, respectively. This is, to our knowledge, one of the first works using magnetically active isotopes to fingerprint the presence of carbon in spin-active quantum emitters. Another interesting possibility is the oxygen-based defect complex~\cite{Li2022b} which seems to match well the spectral profiles of quantum emitters generated by focused ion beams~\cite{Ziegler2019}. The complex $\mathrm{V_BO_N}$ (\figref[b]{microscopic}) is supposed to be paramagnetic with an emission of the ZPL at \qty{2}{eV}. ODMR and EPR measurements could be used to unambiguously identify these oxygen-based emitters.

We would also like to point to a recent work suggesting that tape and organic solvent residues, which are commonly introduced during the exfoliation process of hBN, may lead to the formation of polycyclic aromatic hydrocarbons (PAHs)~\cite{Neumann2023a}. In particular, during the so-called activation step at high temperature and in an inert atmosphere, such as argon, organic residues trapped below the hBN flakes can react to form PAH molecules. These are known to possess excellent quantum optical properties~\cite{Toninelli2021}. The challenge here is that spectra from PAHs show features similar to crystallographic defects in hBN, with an identifiable ZPL followed by a phonon sideband at 160-170~meV. Previous studies on CVD-grown hBN demonstrated that oxygen plays a key role in the photobleaching of such molecular emitters~\cite{Li2023}, similar to laser-induced photobleaching of dye molecules~\cite{Eggeling1998}. Further research involving advanced cleaning processes to ensure the removal of all organic residues would be needed to clearly identify the contribution of PAH and crystallographic defects in the generation of \qty{2}{eV} emitters.

The most promising quantum emitters generated in hBN in terms of quantum optical properties, such as photon purity $\mathrm{g^{(2)}(0)}$ and coherence, are the B-centers~\cite{Shevitski2019, Fournier2021, Gale2022}. They can be generated on demand and with high spatial resolution using keV electron beams. B-centers display a ZPL energy around \qty{2.85}{eV} ($\approx$ 435~nm) with a linewidth on the order of a GHz~\cite{Horder2022, Horder2025}, which is even close to lifetime-limited on timescales shorter than their spectral diffusion~\cite{Gerard2025}. They can be reproducibly created with the same ZPL energy and they possess good quantum optical properties,~\cite{Horder2022} with a high purity as well as a good quantum coherence~\cite{Fournier2023, Horder2025, Gerard2025}. Cathodoluminescence investigations seem to indicate that the generation of B-centers is correlated with the presence of emission at \qty{4.1}{eV}~\cite{Gale2022}, as observed in previous works~\cite{Taniguchi2007, Bourrellier2016, Pelini2019, Vuong2016}. The emission at \qty{4.1}{eV} could correspond to the presence of carbon impurities~\cite{Taniguchi2007, Pelini2019} or to the $\mathrm{C_B C_N}$ carbon dimer~\cite{Jara2021, Mackoit-Sinkeviciene2019, Winter2021, Auburger2021}. The exact formation mechanism of the B-centers is still unknown, but electrons with tens of keV can lead to radiolysis. The ionized electrons and radicals can react to form new defect complexes (see also \secref{e-beam}). Under the action of the electron beam, $\mathrm{C_BC_N}$ defect complexes present in the hBN lattice could react with interstitial boron atoms~\cite{Zhigulin2023a} and lead to the formation of the B-centers. A more recent work~\cite{Nedic2024} points towards electron beam-induced electro-migration of charged defects clustering together with native vacancies and interstitial atoms to form the B-center. 
Polarization-dependent measurements showed that the polarization of the emission is preferentially oriented along directions separated by \qty{60}{\degree}, which could correspond to the crystallographic orientations of the hBN lattice~\cite{Horder2024}. Regarding the microscopic nature of the B-centers, a recent investigation found a nonlinear Stark effect and the absence of a permanent dipole for these centers~\cite{Zhigulin2023a}. Due to symmetry arguments, this observation seems to indicate that the B-centers are consistent with split interstitial configurations~\cite{Ganyecz2024a} (\figref[i]{microscopic}). The formation of the B-centers has been related to the presence of the carbon-based emitters $\mathrm{C_BC_N}$ at \qty{4.1}{eV}, so the split interstitial carbon structure seems to be a good candidate for this color center~\cite{Zhigulin2023a}. Even more recently, the same group generated B-centers in hBN and rhombohedral boron nitride (rBN) with a keV electron beam~\cite{Gale2025}. The blueshift observed experimentally between the ZPL of these emitters in hBN and those in rBN are well captured by DFT calculations~\cite{Gale2025, Maciaszek2024} and hints towards a newly proposed defect complex, the \emph{trans}-C4 configurations. Another research group created blue emitters~\cite{Liang2023c} by keV Helium ion irradiations, but, in contrast to prior works on B-centers, they observed a new class of blue emitters with ZPL centered at \qty{2.73}{eV}. Based on TEM imaging of the e-beam irradiated samples, the study speculates that the emission around \qty{2.73}{eV} is related to boron interstitial $\mathrm{B_i}$ intercalated between the hBN layers, which are sketched in \figref[j]{microscopic} along with other possible interstitial configuration for multilayer hBN.\\

\begin{table*}[tphb]
  \centering
  \begin{tabular}{>{\centering\arraybackslash}p{2.5cm} >{\centering\arraybackslash}p{2.cm} >{\centering\arraybackslash}p{1.5cm} >{\centering\arraybackslash}p{2.5cm} >{\centering\arraybackslash}p{2cm} >{\centering\arraybackslash}p{1.5cm} >{\centering\arraybackslash}p{1.5cm} >{\centering\arraybackslash}p{2cm}}
  \toprule
    
    \multirow{2}{*}{Method} & \multirow{2}{*}{Material} & \multirow{2}{*}{Refs.} & \multirow{2}{*}{$\mathrm{E_{ZPL}}$ [eV]} & \multirow{2}{*}{FWHM [GHz]} & Lifetime [ns] & \multirow{2}{*}{$g^{(2)}(0)$} & \multirow{2}{*}{g-factor} \\  
     \cmidrule(r){1-8}

    \multirow{3}{*}{Strain (intrinsic)} & 1L MoSe\textsubscript{2} & \cite{Branny2016} & 1.58-1.62 &36-97 at 4 K&N.A.&N.A.& 3.8 \\
    \cmidrule(r){2-8}
             & \multirow{2}{*}{1L WSe\textsubscript{2}} & \multirow{2}{*}{\cite{He2015, Koperski2015, Chakraborty2015, Tonndorf2015, Srivastava2015}} & \multirow{2}{*}{$\approx$1.65-1.71} & \multirow{2}{*}{$\approx$30} & \multirow{2}{*}{0.5-7} & 0.03-0.2 at \qty{4}{K} & \multirow{2}{*}{7-12} \\
    \cmidrule(r){1-8}

    \multirow{4}{*}{Strain pillar (SiO$_2$)} & \multirow{2}{*}{1L WSe\textsubscript{2}} & \multirow{2}{*}{\cite{Palacios-Berraquero2017, Wu2025a}} & \multirow{2}{*}{$\approx$1.51-1.7} & \multirow{2}{*}{$\approx$43} & \multirow{2}{*}{3-12} & 0.01 (\qty{4}{K}) - 0.09 (\qty{10}{K}) &\multirow{2}{*}{N.A.}\\
    \cmidrule(r){2-8}
    & \multirow{2}{*}{ML GaSe} & \multirow{2}{*}{\cite{Luo2023a}} & \multirow{2}{*}{1.65-2.06} & \multirow{2}{*}{400-2000} & \multirow{2}{*}{4-6} & 0.3-0.5 at 3.5 K &\multirow{2}{*}{N.A.}\\
    \cmidrule(r){1-8}

    Strain pillar (Au) & 1L WSe\textsubscript{2} & \cite{Cai2018} & $\approx$1.59-1.7 & 240 & 0.7-3.7 & 0.3 at \qty{3}{K} & N.A.\\
    \cmidrule(r){1-8}

    Strain (electrostatic) & \multirow{2}{*}{1L WSe\textsubscript{2}} & \multirow{2}{*}{\cite{Wu2025b}} & \multirow{2}{*}{1.57-1.67} & \multirow{2}{*}{220} & \multirow{2}{*}{2.6} & \multirow{2}{*}{0.4 at \qty{4}{K}} & \multirow{2}{*}{N.A.}\\
    \cmidrule(r){1-8}

    Strain gap (dielectric) & \multirow{2}{*}{1L WSe\textsubscript{2}} & \multirow{2}{*}{\cite{Sortino2021}} & \multirow{2}{*}{$\approx$1.6-1.67} & \multirow{2}{*}{110} & \multirow{2}{*}{$\approx$10-100}& \multirow{2}{*}{0.26 at \qty{4}{K}} &\multirow{2}{*}{N.A.}\\
    \cmidrule(r){1-8}
    
    Strain dome (H-implantation)& \multirow{2}{*}{1L WS\textsubscript{2}} & \multirow{2}{*}{\cite{Cianci2023}} & \multirow{2}{*}{$\approx$1.9-1.97} & \multirow{2}{*}{$\approx$200-1200} & \multirow{2}{*}{3-7} & 0.15-0.25 at \qty{7}{K} & \multirow{2}{*}{7-9}\\
    \cmidrule(r){1-8}

    Strain dent (molding) & \multirow{2}{*}{1L MoSe\textsubscript{2}} & \multirow{2}{*}{\cite{Yu2021}} & \multirow{2}{*}{$\approx$1.58} & \multirow{2}{*}{40} & \multirow{2}{*}{0.2} & \multirow{2}{*}{0.3 at \qty{1.8}{K}} & \multirow{2}{*}{4.1} \\    
    \cmidrule(r){1-8}
    
    \multirow{4}{*}{Strain and e-beam} & \multirow{2}{*}{1L WSe\textsubscript{2}} & \multirow{2}{*}{\cite{Xu2022}} & \multirow{2}{*}{1.55-1.7} &\multirow{2}{*}{N.A.}& \multirow{2}{*}{N.A.} & $\approx$0.3 at \qty{3.5}{K} &\multirow{2}{*}{N.A.}\\
    \cmidrule(r){2-8}
     & \multirow{2}{*}{1L WSe\textsubscript{2}} & \multirow{2}{*}{\cite{Parto2021}} & \multirow{2}{*}{1.52-1.58} & \multirow{2}{*}{20} & \multirow{2}{*}{4-6} & 0.27 at \qty{150}{K} &\multirow{2}{*}{N.A.} \\
    \cmidrule(r){1-8}
    
    & \multirow{2}{*}{2L WSe\textsubscript{2}}& \multirow{2}{*}{\cite{Branny2017}} & \multirow{2}{*}{1.48-1.59} & \multirow{2}{*}{$\approx$30} & \multirow{2}{*}{4.8} & 0.03 at \qty{3.5}{K} & \multirow{2}{*}{N.A.}\\
     \cmidrule(r){2-8}
    \multirow{2}{*}{Strain pillar (resist)}& \multirow{2}{*}{ML InSe} & \multirow{2}{*}{\cite{Zhao2025}} & \multirow{2}{*}{0.80-0.95} & \multirow{2}{*}{$\approx$650} & \multirow{2}{*}{3-90} & 0.34-0.6 at \qty{10}{K} &\multirow{2}{*}{N.A.}\\
    \cmidrule(r){2-8}
    & \multirow{2}{*}{ML MoTe\textsubscript{2}} & \multirow{2}{*}{\cite{Zhao2021}} & \multirow{2}{*}{0.8-1.1} & \multirow{2}{*}{800} & \multirow{2}{*}{22-160} & 0.06-0.18 at \qty{11}{K}& \multirow{2}{*}{3.6} \\
    \cmidrule(r){1-8}

    & 1L WSe\textsubscript{2} & \cite{Abramov2023} & 1.55-1.72 & 600 & 1-30 & 0.02-0.15 at \qty{7}{K} &N.A.\\
    \cmidrule(r){2-8}
    Strain (AFM-indentation) & \multirow{2}{*}{1L WSe\textsubscript{2}} & \multirow{2}{*}{\cite{Rosenberger2019}} & \multirow{2}{*}{$\approx$1.61-1.7} & \multirow{2}{*}{$\approx$150} & \multirow{2}{*}{8} & 0.33-0.41 up to \qty{60}{K} &\multirow{2}{*}{N.A.} \\
     \cmidrule(r){2-8}
    & \multirow{2}{*}{1L WSe\textsubscript{2}} & \multirow{2}{*}{\cite{So2021a}} & \multirow{2}{*}{1.58-1.69} & \multirow{2}{*}{$\approx$100} & \multirow{2}{*}{$\approx$2-8} & 0.3-0.4 at \qty{4}{K} & \multirow{2}{*}{N.A.}\\
     \cmidrule(r){1-8}

    Low-energy e-beam & \multirow{2}{*}{1L MoS\textsubscript{2}} & \multirow{2}{*}{\cite{Dash2025}} & \multirow{2}{*}{1.77-1.85} &\multirow{2}{*}{>170} & \multirow{2}{*}{N.A.} & \multirow{2}{*}{N.A.} &\multirow{2}{*}{0.95}\\
    \cmidrule(r){1-8}

     \multirow{2}{*}{UV irradiation} & \multirow{2}{*}{1L MoS\textsubscript{2}} & \multirow{2}{*}{\cite{Wang2022}} & \multirow{2}{*}{1.75-1.8} &\multirow{2}{*}{>550} & \multirow{2}{*}{100-140} & 0.05-0.42 at \qty{5}{K} &\multirow{2}{*}{N.A.}\\
    \cmidrule(r){1-8}

    & 1L MoSe\textsubscript{2} & \cite{Chen2023b} & 1.56 & 4000 & 10-20 & ensemble &N.A.\\
    \cmidrule(r){2-8}
    Proton irradiation &1L MoSe\textsubscript{2}, MoS\textsubscript{2}, WSe\textsubscript{2}, WS\textsubscript{2} & \multirow{3}{*}{\cite{Zhang2022a}} & \multirow{3}{*}{1.5-1.95} &\multirow{3}{*}{5800-9700} & \multirow{3}{*}{2-200} & \multirow{3}{*}{ensemble} &\multirow{3}{*}{N.A.}\\
    \cmidrule(r){1-8}

    \multirow{2}{*}{Cr-implantation} & \multirow{2}{*}{1L MoSe\textsubscript{2}} & \multirow{2}{*}{\cite{Bui2023}} & \multirow{2}{*}{1.48-1.52} & \multirow{2}{*}{3500} & \multirow{2}{*}{14} & ensemble at \qty{10}{K} & \multirow{2}{*}{1.2} \\
    \cmidrule(r){1-8}
     
     \multirow{2}{*}{He-ion irr.} & \multirow{2}{*}{1L MoS\textsubscript{2}} & \multirow{2}{*}{\cite{Klein2019, Klein2021a, Hotger2023a, Hoetger2023, Barthelmi2024}} & \multirow{2}{*}{1.75-1.85} & \multirow{2}{*}{60} & \multirow{2}{*}{270-1730} & 0.23-0.44 at \qty{1}{K} & \multirow{2}{*}{0.1-1.3} \\
    \cmidrule(r){1-8}
    
     Nb-doping & 1L WS\textsubscript{2} & \cite{Loh2024} & 1.92-1.93 & 75 & 80 & 0.27 at \qty{4}{K} & 10.4 \\

    \bottomrule
  \end{tabular}
  \caption{Table gathering photophysical information about quantum emitters created by various methods in TMDCs and related materials. g-factors are quoted as absolute values. 1L: single layer, ML: multilayer, N.A.: not available.}
  \label{tab:tmdcs}
\end{table*} 

\emph{For TMDCs}. In contrast to hBN, SPEs in 2D TMDCs exhibit rather low detrapping barriers (on the order of tens of meV) and require consequently cryogenic temperatures to be optically active. Their microscopic origin is mostly interpreted in terms of strain-localized, impurity-localized, or vacancy-localized excitons, although color centers based on deep levels with emission energies in the telecom range have been predicted for some of the large gap TMDCs \cite{Gupta2019} as well. \tabref{tmdcs} summarizes SPEs (and some other defect-associated emitters) in the most common TMDCs (MoS\textsubscript{2}, MoSe\textsubscript{2}, WSe\textsubscript{2}, WS\textsubscript{2}, MoTe\textsubscript{2}) and some monochalcogenides (GaSe and InSe). Towards narrowing down the origin of the single-photon emission particularly useful experimental fingerprints are the doping dependence and g-factor of the defect-related excitons. In strained WSe\textsubscript{2} and WS\textsubscript{2}, large g-factors ($|g|\approx 8 - 10 $) are found, and they are associated with dark exciton states. In turn, the single-photon emission is attributed to strain funneling and localization of dark excitons\cite{Rosati2021}, which hybridize with shallow trap states close to the conduction band (cf. \figref[d]{defect_models}) \cite{Linhart2019}. Due to strong localization at the defect sites, the optical selection rules are relaxed resulting in an effective brightening and radiative recombination of the dark states. Note that the experiments are consistent with an intrinsic origin of the defect states, such as shallow impurities or vacancies, but future studies might reveal the impact of possible extrinsic effects from e.g. absorbed and chemisorbed molecules as well (cf. \secref{e-beam}). For strain-induced SPEs in MoSe\textsubscript{2} and MoTe\textsubscript{2}, g-factors consistent with a bright exciton state were found ($|g|\approx 4$). In the case of unstrained 2D TMDCs with impurities, for example Cr-implanted MoSe\textsubscript{2} and Nb-doped WS\textsubscript{2}, the predominant origin of exciton localization appears to be donor/acceptor-bound excitons (cf. \figref[b]{defect_models}). Nevertheless, the measured g-factors can vary significantly from $|g|\approx 1$ for Cr-implanted MoSe\textsubscript{2}\cite{Bui2023} to $|g|\approx 10$ for Nb-doped WS\textsubscript{2}\cite{Loh2024}, which generally point to different admixtures of localized states and extended states in the radiative recombination process \cite{Amit2022}. For Nb-WS\textsubscript{2}, the exclusive incorporation of Nb on W-sites, as would intuitively be expected for a dilute metallic dopant, was verified by STEM enabling a detailed comparison of DFT-calculated electronic structure and single particle g-factors. Calculations place the Nb-induced defect bands just below the valence band maximum resulting in significant defect-to-band hybridization. Based on the large g-factor, a likely scenario is a defect-induced brightening of a spin-forbidden transition involving a valence band state. Therefore, decay of neutral dark excitons bound to ionized Nb, denoted as Nb$^-$X$^0_\text{D}$, was proposed as the dominant radiative recombination pathway leading to the observed SPE with $g \approx 10$. By contrast, for Cr-implantation of MoSe\textsubscript{2}\cite{Bui2023} several different defect geometries, including Cr adatoms, Cr on Mo-sites, and Cr at Se-sites with a Se adatom, are considered likely based on molecular dynamics simulations. In turn, an assignment of the optical transition based on the observed g-factor alone is difficult. Due to the absence of the defect emission for p-type doping in field effect devices, radiative recombination of an electron bound to a defect state with a valence band hole was tentatively proposed. The small negative g-factor of $g\approx -1.2$ may potentially be explained by a large g-factor for the electron acceptor level (compensating the large g-factor of holes near the valence band maximum), as would be case for Cr on a Mo-site, or by a reduced g-factor of holes in the valence band due hybridization with the defect, as would be the case for Cr on a Se-Site. For proton or He-ion beam irradiation, experiments point towards a predominant creation of chalcogen vacancies \cite{Chen2023b,Mitterreiter2020,Mitterreiter2021} narrowing the discussion and theoretical modeling to these structures. For SPEs in He-ion exposed MoS\textsubscript{2}, a vanishing valley polarization and a small positive g-factor (between +0.1 and +1.3) were measured experimentally \cite{Hoetger2023}. Both observations were found to be consistent with exciton transitions where the presence of sulfur vacancies results in strong hybridization between defect and band states (cf. \figref[c]{defect_models}) as calculated by BSE-GW calculations\cite{Refaely-Abramson2018, Amit2022}. Moreover, the predicted optical absorption characteristics are consistent with photovoltage spectra as detected in corresponding vertical tunneling devices \cite{Hotger2023a}. For proton irradiated MoSe\textsubscript{2}\cite{Chen2023b}, a finite valley polarization was observed for the defect emitters. Therefore, band excitons which retain the valley character bound to deep acceptor levels introduced by selenium vacancies (cf. \figref[b]{defect_models}) were proposed as the microscopic origin. The diversity of observation and interpretations for SPEs in both hBN and TMDCs highlight the importance of an unambiguous atomic scale characterization, which we discuss in the next section.

\section{Towards atomistic control and characterization}
\label{sec:atomistic}

\begin{figure*}[thb]
  \centering
  \includegraphics[width=1.95\columnwidth]{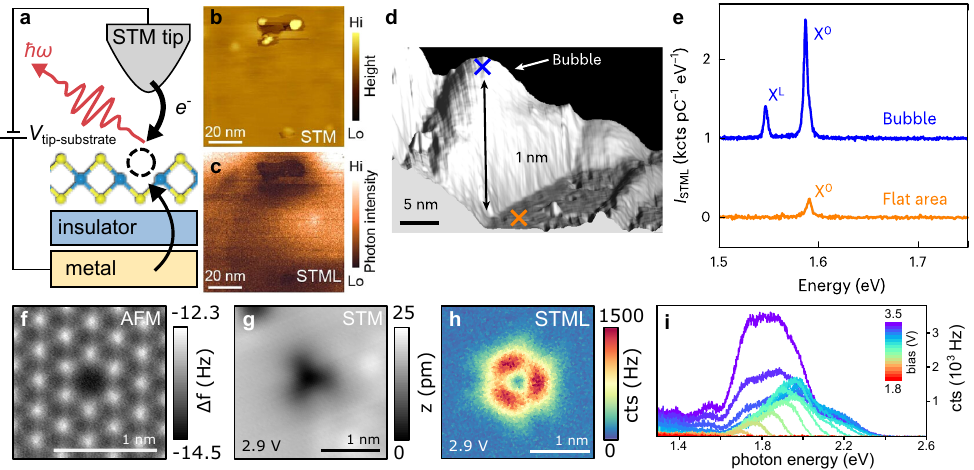}
  \caption{\textbf{Atomic scale characterization of defect-based SPEs.} \textbf{a} Schematic of scanning tunneling microscope induced luminescence (STML). \textbf{b} STM topography of a WSe\textsubscript{2} monolayer with a hBN decoupling layer. \textbf{c}  Topographic features correlate with nanoscale variations in STML. Reproduced with permission from \citet{geng2024imaging} ACS Nano \textbf{18}, 8961-8970 (2024). Copyright 2024 American Chemical Society. \textbf{d} STM topography of a MoSe\textsubscript{2} monolayer with a graphite decoupling layer on Au(111). \textbf{e} The graphite provides sufficient decoupling to acquire STML spectra resolving both free $X_0$ and localized excitons $X_\text{L}$ with nanoscale spatial resolution. Reproduced with permission from \citet{Lopez2023} Nature Materials \textbf{22}, 482–488 (2023). Copyright 2023, The Author(s), under exclusive license to Springer Nature Limited. \textbf{f} Non-contact AFM image of monolayer WS\textsubscript{2} directly placed on epitaxial graphene. The bright features denote the positions of the sulfur atoms with a single sulfur vacancy at the center of the image. \textbf{g} STM image revealing the defect orbital of the sulfur vacancy. The tip-substrate voltage is \qty{2.9}{V}. \textbf{h} STML mapping of a single sulfur vacancy. The tip-substrate voltage is again \qty{2.9}{V}. \textbf{i} Bias-dependent luminescence spectra. Due to the missing decoupling layer to the substrate, a fast plasmon-mediated emission process prevails over the slow recombination dynamics of the defect-localized exciton. Reproduced with permission from \citet{schuler2020electrically}, Sci. Adv. \textbf{6}, eabb5988 (2020). Copyright 2020 Authors, licensed under a Creative Commons Attribution Non Commercial 4.0 (CC BY-NC) License.}
  \label{fig:atomistic}
\end{figure*}

An emerging technique to correlate atomic structure and optical properties of defect-based SPEs is scanning tunneling microscope luminescence (STML) as sketched in \figref[a]{atomistic}. Electrons tunnel from the STM tip into local states in the 2D layer. In a simple picture, if the electrons are injected with sufficiently high excess energy, which is controlled by the tunneling bias $V_\text{tip-sample}$, they can relax by emitting photons, which are then detected in the far field~\cite{gutzler2021light, roslawska2020atomic}. Following early works from around 1990~\cite{Coombs1988a, Berndt1993}, STML has achieved so far spatial imaging with atomic resolution~\cite{nazin2003atomic}, vibronic coupling~\cite{Qiu2003, kong2021probing}, time-resolved luminescence~\cite{Dolezal2024}, and single-photon statistics~\cite{zhang2017electrically,leon2020single}. A key challenge is detecting emission spectra which are specific to the molecular or defect structure in which electrons are being tunneled, because the junction between the metallic tip and the metal substrate forms a plasmonic cavity~\cite{Schaeverbeke2019}, which can dominate the light emission process and spectral fingerprints. To overcome this limitation, a decoupling layer needs to be introduced between the localized state and the metal substrate (\figref[a]{atomistic}). In a simplified picture, the decoupling layer extends the lifetime of the excited electronic state in the molecule~\cite{kaiser2023charge} or defect~\cite{Bobzien2025} by preventing fast tunneling into or quenching by the metal substrate, which in turn allows the sample-specific photon emission to prevail. For molecules, few-layer NaCl evaporated in-situ is a common barrier layer~\cite{kaiser2023charge}. For 2D TMDCs, hBN barriers can decouple the 2D layer and the defects therein from the metallic substrate, such as graphene~\cite{nisi2024scanning} or Au~\cite{geng2024imaging}. 
\figref[b, c]{atomistic} compare topography and STM-induced electroluminescence of a WSe\textsubscript{2} monolayer decoupled from the substrate by a 10~nm thick hBN layer and connected in a lateral fashion to a gold substrate. 
The spatial maps of the exciton electroluminescence exhibit nanoscale variations that correlate with the STM topography. Interestingly, efficient excitonic STML and nanoscale imaging were also achieved in monolayer MoSe\textsubscript{2} that was only weakly decoupled by few-layer graphite from a Au~(111) substrate~\cite{Lopez2023} (\figref[d, e]{atomistic}). This may be related to the strong oscillator strength and concomitant short radiative lifetime of free excitons in 2D TMDCs. The spectra clearly differentiate tip-induced luminescence from free excitons and excitons localized to nanoscale topographic features, such as bubbles. As a possible excitation mechanism for the electroluminescence, a many-body excited-state picture was proposed~\cite{Miwa2019, Lopez2023}. Electrons (holes) injected into the conduction (valence) band form a negatively (positively) charged excited-state resonance. The system can either directly relax back to the neutral ground state, where the electron (hole) tunnels into the substrate, or it can form an intermediate exciton state by binding a hole (electron) tunneled into the 2D layer from the substrate. Note that, due to the Coulomb interaction, the exciton state represents an intermediate energetic minimum. Now, the system can relax to the neutral ground state by emission of a photon. This picture highlights the importance of balancing the tunneling rates between tip-sample, for example by controlling the tip-sample and the tip-substrate distance, for example by introducing a suitable decoupling layer. Towards the ultimate goal of identifying and controlling single-photon sources in 2D vdW materials by STML, \figref[f-i]{atomistic} depicts experiments on single sulfur vacancies in monolayer WS\textsubscript{2} on epitaxial graphene~\cite{schuler2020electrically}. Non-contact atomic force microscopy using a CO-functionalized tip resolves a single lattice defect at a sulfur site (\figref[f]{atomistic}). The electronic spectroscopy by STM identifies the defect as a sulfur vacancy (\figref[g]{atomistic}). Mapping of the STML demonstrates efficient driving of electroluminescence by tunneling into the defect orbital (\figref[g]{atomistic}). However, the photon spectra indicate a dominant plasmonic light emission process, which is likely due to the absence of a decoupling layer and a rather long radiative lifetime, as measured for SPEs in MoS\textsubscript{2} \cite{Klein2021a}. Despite the absence of a clear spectral fingerprint and of single-photon statistics, the latter examples highlight the progress towards a complete atomistic understanding of SPEs in 2D materials. 

Similarly, significant progress towards correlating atomic structure and optical properties via (transmission) electron microscopy has been reported. For example, excitonic wavefunctions in an atomically thin heterostructure were reconstructed by precise low-energy electron loss spectroscopy~\cite{susarla2022hyperspectral}. Using STEM-based CL, multiple classes of stable quantum emitters were resolved in hBN~\cite{Hayee2020}. The spatial resolution was demonstrated to be about 15~nm. While this still precludes the study of single defects with sub-nm precision, as discussed above for STM, it is sufficient to elucidate, for example, the role of nanoscale strain patterns on the optical properties of the quantum emitters~\cite{Hayee2020}. For TMDCs, a recent work employed transmission electron microscopy combined with cathodoluminescence to generate, identify and optically address sulfur mono- and bivacancy defects in monolayer WS\textsubscript{2}, although single-photon emission was not resolved~\cite{Sun2024} in CL. Due to the often complicated interplay of defect generation, imaging, and excitation via the electron beam, machine learning and artificial intelligence-based algorithms are being increasingly utilized to identify atomic scale defects and to study their formation under the electron beam~\cite{ Ziatdinov2017,Lee2020, Yang2021,Roccapriore2023, Sun2024,Weile2025}. Recent advancements in algorithmic electron-beam control have enabled the systematic targeting of individual atoms in monolayer TMDCs with few-picometer precision, facilitating the controlled creation vacancies~\cite{Roccapriore2025}. This was achieved through a sparse, rapid, and dose-efficient scan combined with a fast algorithm (“atomic lock-on”), which together enable rapid lattice reconstruction while mitigating common issues such as scan distortions and sample drift. 

\section{Conclusion and Outlook}
\label{sec:outlook}

\begin{figure*}[thb]
  \centering
  \includegraphics[width=1.95\columnwidth]{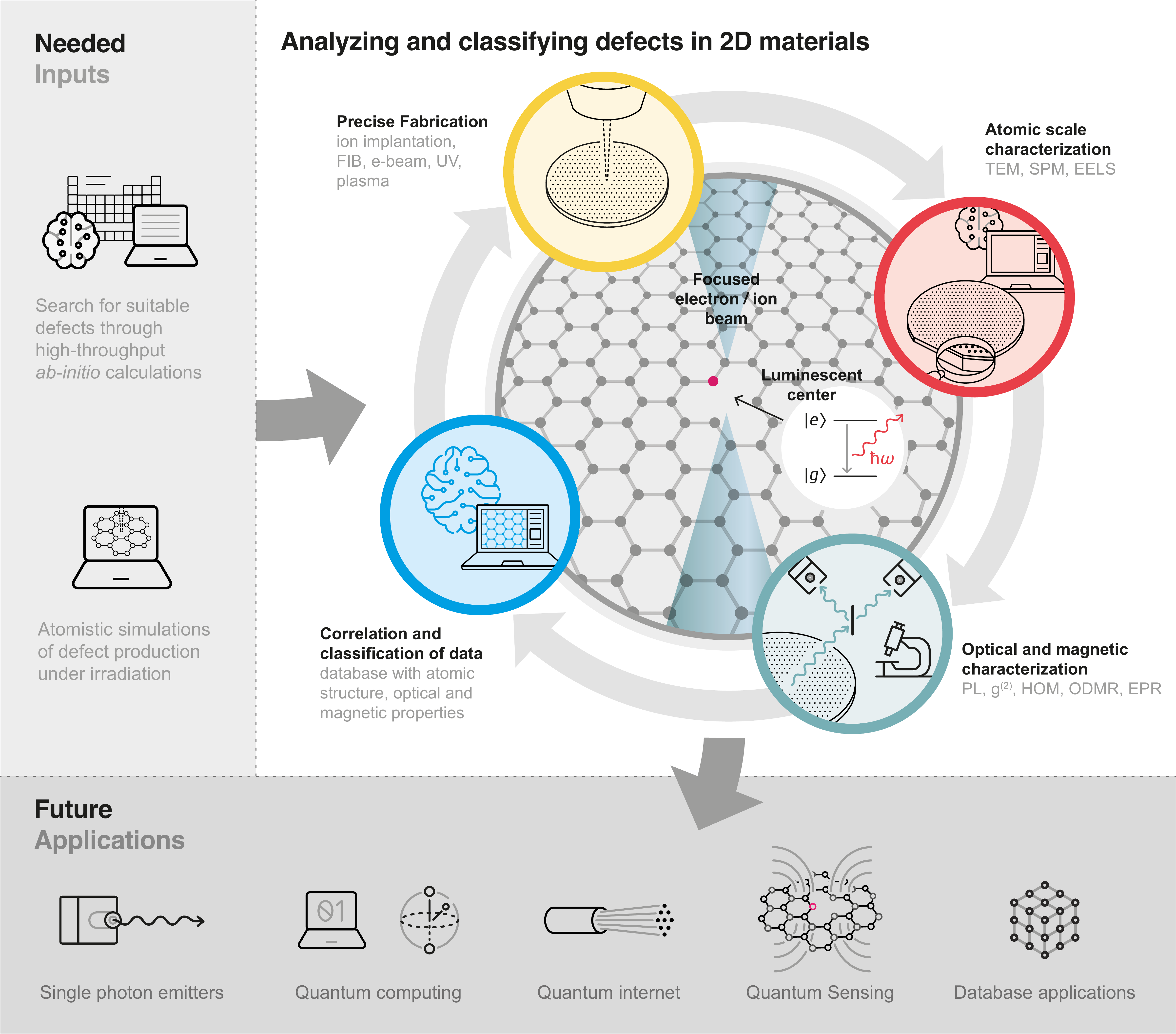}
  \caption{\textbf{Towards a systematic and predictive Framework.} Schematic illustration of a theory-guided strategy for the controlled generation of quantum emitters in 2D materials like hBN and TMDCs for future quantum applications. FIB: focused ion beam. UV: ultraviolet. TEM: transmission electron microscope. SPM: scanning probe microscopy. EELS: electron-energy loss spectroscopy. PL: photoluminescence. $g^{(2)}$: second-order correlation function for photon statistics measurements. HOM: Hong-Ou Mandel setup for photon indistinguishability measurements. ODMR: optically detected magnetic resonance. EPR: electron paramagnetic resonance.}
  \label{fig:outlook}
\end{figure*}

In this review, we provided a broad overview of fabrication methods used to create quantum emitters in TMDCs and hBN. We covered various approaches, including strain engineering (\secref{strain}), low-energy ion irradiation (\secref{plasma}), high-energy irradiation methods (\secref{ion}), ion implantation (\secref{implantation}), electron beam irradiation (\secref{e-beam}), and photon-based irradiation (\secref{xray}). From a fundamental physics point of view, a key challenge remains the identification and understanding of the microscopic origin of the single-photon emission, which we discussed extensively in \secref{nature} and \secref{atomistic}.

Irradiation techniques with charged particles such as ions or electrons offer nanometer-scale probe resolution, making them key tools for the site-selective creation of quantum emitters in 2D materials. These methods have successfully produced several classes of emitters in hBN — including those emitting in the near-infrared (NIR), around \qty{2}{eV}, blue, and ultraviolet spectral regions — and predominantly NIR-emitting centers in monolayer TMDCs. 
In hBN, electron beam irradiation has been particularly effective in generating blue emitters with reproducible ZPL energies and precise spatial localization. Both high- and low-energy ion irradiation methods have consistently demonstrated the ability to create ensembles of magnetically active boron vacancy {\VB} centers. However, the site-controlled creation of quantum emitters with ZPLs around \qty{2}{eV} remains a significant challenge. These \qty{2}{eV} emitters are particularly promising due to their narrow ZPLs and well-defined phonon sidebands. Emitters formed through high-temperature annealing have even approached lifetime-limited linewidths — an essential requirement for indistinguishable single-photon sources \cite{Dietrich2018, Dietrich2020, Akbari2022}. 
Lifetime-limited quantum emitters at \mbox{\qty{2}{eV}} have not yet been demonstrated using irradiation or strain-based techniques.

Oxygen plasma~\cite{Exarhos2019} and $^{13}\mathrm{CO}_2$ implantation~\cite{Gao2025} revealed the existence of magnetically active emitters in the \qty{2}{eV} range. More studies in this direction are needed to create spin-active \qty{2}{eV} emitters on demand and with a higher yield for potential sensing and quantum memory applications. These efforts should be paired with advanced cleaning protocols, such as those from \citet{Neumann2023a}, to mitigate the formation of spurious molecular emitters arising from organic residues on hBN flakes during the different fabrication stages\cite{Gasparutti2020}. Recent theoretical\cite{Wang2025} and experimental\cite{Smit2023} studies have further indicated that 2D materials can serve as ideal hosts for molecule-based quantum emitters, highlighting the importance of surface cleanliness in order to differentiate crystallographic defects from molecular emitters. In the TMDC family, strain-induced single-photon emitters (SPEs) currently represent the most advanced and widely adopted platform. Ion beam engineering of SPEs has been particularly successful in MoS$_2$ \cite{Barthelmi2020}, with promising progress now extending to other TMDCs \cite{Zhang2022a, Chen2023b}. Low-energy ion implantation (from tens to hundreds of eV) has also emerged as a promising technique, especially for introducing impurity- and impurity-vacancy complexes \cite{Gupta2019, Thomas2024}, with initial experimental demonstrations underway \cite{Bui2023}.

Potential applications would benefit from further refinements of the spatial and spectral precision of the emitter creation process. For example, while strain-induced emitters in WSe$_2$ are created quite consistently and with high yield in the NIR spectral range, the ZPL of different emitters still exhibits statistical inhomogeneous broadening of about \qty{30}{meV}. Regarding the positioning accuracy of strain-induced emitters, Ref. \onlinecite{Branny2017} reports a direct correlation between the location of the patterned pillar and the observed quantum emission to better than \qty{120}{nm}, while other studies often quote the size of pillars or strain indents as estimates of the positioning accuracy (typical values are \qty{300}{nm} \cite{Parto2021, So2021a}). By contrast, atomically resolved studies of focused He-ion beam irradiated MoS\textsubscript{2} demonstrated a spatial precision better than \qty{9}{nm} for the creation of single point defects, yet without a direct correlation to the location of optical emission \cite{Mitterreiter2020}. The optically active single photon emitters created in the latter way are consistently produced, also using other methods\cite{Wang2022, Dash2025}, but a statistical broadening within ensembles of such emitters of about \qty{30}{meV} is still present \cite{Barthelmi2020, Klein2021a}. With respect to the reproducibility of the ZPL, blue emitters at \qty{435}{nm} (\qty{2.85}{eV}) created by electron beam irradiation of hBN exhibit the most consistent results so far~\cite{Fournier2021, Gale2022}. Spatial positioning of these emitters is further verified to be better than \qty{300}{nm}~\cite{Fournier2021}. Notably, a very recent work employing a focused He-ion beam demonstrated a very narrow statistical distribution of ZPL for a potentially different class of blue emitters at \qty{433}{nm} (\qty{2.86}{eV}) and \qty{454}{nm} (\qty{2.73}{eV}) with an inhomogeneous broadening as low as \qty{0.1}{nm} (\qty{0.7}{meV})~\cite{Liang2023c}.

\emph{Outlook and Future Directions} Future research should address the critical issue of coherence loss, particularly due to phonon interactions \cite{Preuss2022, Fischer2023, Piccinini2024}. Non-perturbative treatments of electron-phonon coupling \cite{Nazir2016, Groll2021, Svendsen2022} are essential to enable phonon suppression schemes, such as the swing-up of quantum emitter population (SUPER) excitation scheme \cite{Bracht2021, Karli2022, Vannucci2024}. Moreover, the development of quantum emitters operating at telecom wavelengths remains a priority for integration into fiber-based quantum networks. While strain-induced emitters in MoTe$_2$ have demonstrated telecom emission \cite{Zhao2021}, recent work has also reported single-photon emission from strained InSe multilayers \cite{Zhao2025}.
High-throughput theoretical frameworks and curated color center databases \cite{Thomas2024, Cholsuk2024, Bertoldo2022} could significantly accelerate the identification of defect complexes with optimal optical and magnetic properties across the vast configuration space of 2D and vdW layered materials.
Despite rapid experimental progress, the microscopic origin of many quantum emitters in hBN and TMDCs remains poorly understood, as discussed in \secref{nature}. Establishing this foundational knowledge is critical for enabling on-demand, site-selective fabrication of quantum emitters with properties comparable to those in III–V semiconductors. A clear understanding of emitter structure will inform strategies for improving brightness, purity, and indistinguishability, while also enabling precise modeling of environmental coupling and the rational design of pulsed excitation schemes like SUPER.
Special attention must be paid to fabrication artifacts, such as those arising from tape transfer processes and solvent residues. Based on the report by \citet{Neumann2023a}, we advocate for transparent reporting of fabrication protocols, rigorous substrate cleaning, and standardized reference tests to suppress parasitic emitters. On the theoretical front, models incorporating thermal broadening \cite{White2021, Fu2009, Vannucci2024} and phonon-induced decoherence\cite{Iles-Smith2017} will be essential to guide the development of emitters with robust performance at elevated temperatures.

\emph{Towards a Systematic and Predictive Framework}
The controlled creation of quantum emitters in 2D materials via irradiation and strain engineering remains a promising yet largely empirical field. Realizing its full potential will require a transition to systematic, theory-guided strategies~\cite{Bassett2019, Ping2021, Bertoldo2022, Thomas2024, Cholsuk2024}. To date, numerous irradiation approaches — including electron and ion beams — have been deployed to produce emitters in hBN and TMDCs with varying outcomes. Electron beams in the tens of keV range have yielded emitters with near-Fourier-limited linewidths~\cite{Gerard2025, Gerard2025a}, whereas ion beams provide high creation yields and remain the most widely used method.
However, most discoveries have emerged through trial and error. A more predictive methodology is needed, integrating high-throughput \textit{ab-initio} calculations, atomistic simulations, and detailed experimental characterization (see \figref{outlook}). Such an interdisciplinary approach would enable the targeted design of emitters with engineered quantum optical and magnetic features.
Research should expand beyond traditional defects (e.g., carbon, vacancies, and oxygen) to include novel dopants identified by deep-learning-assisted screening. For example, rare-earth elements like erbium could offer telecom emission~\cite{Garcia-Arellano2025} and enhanced coherence through shielded inner-shell transitions. Defects with large zero-field splitting may also enable long-lived spin coherence, a valuable asset for quantum memory applications.
Ion implantation remains the most promising avenue for introducing such engineered defects. Given the ultrathin nature of 2D materials, the implantation process must be carefully optimized — considering kinetic energy, incidence angle, and substrate effects — to ensure dopant incorporation rather than transmission or backscattering (see \secref{fundamentals}). Reactive-gas-assisted focused beams may facilitate site-selective doping, while interstitial implantation represents a largely unexplored but potentially transformative approach with minimal environmental coupling. Implanting magnetic isotopes can offer the possibility to clarify the nature of the elements forming these emitters by changing on purpose their magnetic properties via hyperfine coupling, which can then be compared to predictions from \emph{ab-initio} calculations. 
Molecular dynamics simulations, including thermal effects and defect diffusion, are necessary to refine these techniques. Concurrently, \textit{in situ} atomic-scale characterization — using STEM, STM, possibly supplemented by cathodoluminescence measurements — can correlate structural and optical properties~\cite{Hayee2020, Lopez2023}. Advanced techniques such as energy-filtered EELS and low-energy phonon mapping provide insight into defect composition and vibrational environments \cite{Krivanek2014a, Gadre2022, Kim2023a, Venkatraman2019a}.
Machine learning, particularly convolutional neural networks, offers exciting prospects for identifying luminescent centers \cite{Madsen2018, Propst2024, Yang2021, Weile2025}, especially in low-Z materials such as hBN, where imaging resolution is inherently limited. Once identified, these emitters should undergo comprehensive optical characterization — ideally at the same locations used in structural analysis — requiring STEM-compatible sample holders to facilitate such correlative studies.
Spectroscopic measurements of phonon sidebands should be compared to DFT and open quantum system models, which accurately capture phonon-induced decoherence effects \cite{Preuss2022,Fischer2023, Iles-Smith2017}. Furthermore, second-order correlation and Hong-Ou-Mandel (HOM) interference experiments are essential for quantifying single-photon purity and indistinguishability — key metrics for quantum photonics\cite{Fox2025}. Magnetic characterization via EPR and ODMR complements optical studies by providing insight into ground-state spin configurations and coherence.
A further step, which is currently attracting growing attention in the 2D community, is the integration of luminescent centers in vdW materials into more complex, layered structures. The aim is to leverage the flat nature of 2D systems to fabricate more sophisticated heterostructures \cite{Wang2013} in which the optical response of embedded quantum emitters can be modulated and further investigated upon application of a gate voltage and/or a tunneling/source-drain current \cite{Palacios-Berraquero2016, Akbari2022}. The advantages are multiple. On one hand, by applying a tunneling/source-drain voltage it is possible to control the carrier population and the Fermi level of the material, thus injecting carriers into in-gap states \cite{Chandni2015, Guo2023b, Grzeszczyk2024, Park2024} -- including those that are typically not accessible by usual optical excitation \cite{Guo2023b, White2022} -- with the potential to obtain quantum light emission by means of only voltage/current application, i.e. electroluminescence \cite{Song2021, Grzeszczyk2024, Steiner2024, Park2024}. This strategy provides a pathway to access and explore the photophysics of new hidden, dark states in such materials \cite{White2022, Yu2022a}, provide a tool for probing the properties of already known classes of emitters \cite{Hotger2023a, Fraunie2025} and/or provide a knob for turning on/off emitters deterministically \cite{Hotger2021, White2022, Yu2022a}. On the other hand, embedding of luminescent centers in electrically switchable devices can lead to the production of 2D light-emitting diodes (LEDs) \cite{Palacios-Berraquero2016, Wang2017, Song2021} or on-demand single-photon sources which can be turned on/off by means of current application. Several works have already addressed different aspects of device integration, both for TMDCs \cite{Hotger2021, Hotger2023a} and hBN \cite{Noh2018, Gale2023, Grzeszczyk2024, Steiner2024, Fraunie2025, Zhigulin2025}. Future efforts are expected to focus on advancing fabrication strategies and enhancing device performance under ambient conditions \cite{Turunen2022}, especially concerning minimizing the operation-induced device degradation\cite{Zhigulin2025}.

Finally, to streamline research and application, experimental data should be compiled in an open-access searchable database. Classification by defect type, emission wavelength, magnetic properties, and fabrication method will accelerate the integration of these emitters into scalable quantum photonic platforms. Such databases can also serve as training sets for machine learning models, enabling more accurate prediction, screening, and engineering of quantum emitters in 2D materials.

\section*{Conflict of Interest}
The authors have no conflicts to disclose.

\section*{Author Contributions}
All authors contributed to the writing of the manuscript and revised the overall text.

\begin{acknowledgments}
A.W.H and C.K. thank the International Graduate School of Science and Engineering (TUM-IGSSE, project BrightQuanDTUM), the Deutsche Forschungsgemeinschaft (DFG, German Research Foundation) via the Munich Center for Quantum Science and Technology No. (MCQST)-EXC-2111-390814868 and via the excellence cluster e-conversion No. EXC-2089/1-390776260, and the Munich Quantum Valley K6 as well as the One Munich Strategy Forum - EQAP.
C.K. acknowledges funding by the European Union’s Horizon Europe research and innovation programme under grant No. 101076915 (2DTopS).
N. S. thanks the Novo Nordisk Foundation NERD Programme (project QuDec NNF23OC0082957).
N.S. and M.W. acknowledge the support from the Independent Research Fund Denmark, Natural Sciences (project no. 0135-00403B) and from the Danish National Research Foundation through NanoPhoton - Center for Nanophotonics (project no. DNRF147). 
A.H. acknowledges the NNF Biomag project (NNF21OC0066526) and the Danish National Research Foundation through the center for macroscopic quantum states (bigQ, project no. DNRF0142). 
A.V.K. acknowledges funding from the German Research Foundation (DFG), projects KR 4866/9-1, and the collaborative research center “Chemistry of Synthetic 2D Materials” CRC-1415-417590517. 
D.P.B.F.M. and T.W.H. thank the Independent Research Fund Denmark, DFF-FTP (project no. 4264-00065B)
 
\end{acknowledgments}



\section*{References}
\bibliography{bibliography}

\end{document}

%% file: preamble.tex
\usepackage{graphicx,dcolumn,siunitx,bm,mathptmx,braket,changes}
\usepackage{dcolumn}
\usepackage[export]{adjustbox}
\usepackage[utf8]{inputenc}
\usepackage[T1]{fontenc}
\usepackage{hyperref,booktabs,etoolbox}
\usepackage{multirow}

\usepackage{anyfontsize}  

\DeclareSIUnit\angstrom{\text {Å}}

\newcommand*{\figref}[2][]{%
  \hyperref[{fig:#2}]{%
    Figure~\ref*{fig:#2}%
    \ifx\\#1\\%
    \else
      \,(#1)%
    \fi
  }%
}

\newcommand*{\tabref}[2][]{%
  \hyperref[{tab:#2}]{%
    Table~\ref*{tab:#2}%
    \ifx\\#1\\%
    \else
      \,#1%
    \fi
  }%
}

\newcommand*{\secref}[2][]{%
  \hyperref[{sec:#2}]{%
    Section~\ref*{sec:#2}%
    \ifx\\#1\\%
    \else
      \,#1%
    \fi
  }%
}

\newcommand*{\appref}[2][]{%
  \hyperref[{app:#2}]{%
    Appendix~\ref*{app:#2}%
    \ifx\\#1\\%
    \else
      \,#1%
    \fi
  }%
}

\newcommand*{\equationref}[2][]{%
  \hyperref[{eq:#2}]{%
    Equation~\eqref{eq:#2}%
    \ifx\\#1\\%
    \else
      \,#1%
    \fi
  }%
}

\def\VB{$\mathrm{V_B^-}$}